\documentclass[10pt,american,letterpaper]{article}
\usepackage[T1]{fontenc}
\usepackage[utf8x]{inputenc}
\usepackage{geometry}
\usepackage{xcolor}
\usepackage{babel}
\usepackage{prettyref}
\usepackage{float}
\usepackage{units}
\usepackage{mathtools}
\usepackage{amsmath}
\usepackage{amsthm}
\usepackage{amssymb}
\usepackage{stmaryrd}
\usepackage{graphicx}
\PassOptionsToPackage{normalem}{ulem}
\usepackage{ulem}
\usepackage{microtype}
\usepackage[unicode=true,pdfusetitle,
 bookmarks=true,bookmarksnumbered=false,bookmarksopen=true,bookmarksopenlevel=2,
 breaklinks=false,pdfborder={0 0 0},pdfborderstyle={},backref=false,colorlinks=false]
 {hyperref}
\hypersetup{
 unicode=true, bookmarks=true,bookmarksnumbered=false,bookmarksopen=true, breaklinks=false,pdfborder={0 0 0},backref=false,colorlinks=false}

\makeatletter

\providecommand{\tabularnewline}{\\}
\providecolor{lyxadded}{rgb}{0,0,1}
\providecolor{lyxdeleted}{rgb}{1,0,0}

\DeclareRobustCommand{\lyxsout}[1]{\ifx\\#1\else\sout{#1}\fi}

\usepackage{amsmath,amssymb}
\usepackage{bbm}
\usepackage{cite}
\usepackage{nameref,hyperref}
\usepackage[right]{lineno}

\usepackage[aboveskip=1pt,labelfont=bf,labelsep=period,justification=raggedright,singlelinecheck=off]{caption}

\makeatletter
\renewcommand{\@biblabel}[1]{\quad#1.}
\makeatother

\usepackage{tabularx} 
\usepackage{multirow}
\usepackage{colortbl}
\usepackage{dsfont}

\usepackage{prettyref}
\newrefformat{cap}{\hyperref[#1]{Fig~\ref{#1}}}
\newrefformat{fig}{\hyperref[#1]{Fig~\ref{#1}}}
\newrefformat{tab}{\hyperref[#1]{Table ~\ref{#1}}}
\newrefformat{sec}{\hyperref[#1]{Section~\ref{#1}}}
\newrefformat{sub}{\hyperref[#1]{Section~\ref{#1}}}
\newrefformat{eq}{\hyperref[#1]{Eq~\ref{#1}}}

\newcommand\mytabspace[1]{%
 \parbox[c][\totalheightof{#1}+2\fboxsep+2\fboxrule][c]{\widthof{#1}}{#1}%
}

\DisableLigatures[f]{encoding = *, family = * }
\usepackage[T1]{fontenc}
\usepackage{lmodern}

\date{}

\makeatother

\begin{document}
\begin{flushleft}
{\Large
\textbf\newline{Conditions for wave trains in spiking neural networks} 
}
\newline
\\
Johanna Senk\textsuperscript{1*},
Karol\'{i}na Korvasov\'{a}\textsuperscript{1,2},
Jannis Schuecker\textsuperscript{1},
Espen Hagen\textsuperscript{1,3},
Tom Tetzlaff\textsuperscript{1},
Markus Diesmann\textsuperscript{1,4,5},
Moritz Helias\textsuperscript{1,5}
\\
\bigskip \textbf{1} Institute of Neuroscience and Medicine (INM-6) and Institute for Advanced Simulation (IAS-6) and JARA Institute Brain Structure-Function Relationships (INM-10), Jülich Research Centre, Jülich, Germany
\\
\textbf{2} Faculty 1, RWTH Aachen University, Aachen, Germany
\\
\textbf{3} Department of Physics, University of Oslo, Oslo, Norway 
\\
\textbf{4} Department of Psychiatry, Psychotherapy and Psychosomatics, Medical Faculty, RWTH Aachen University, Aachen, Germany
\\
\textbf{5} Department of Physics, Faculty 1, RWTH Aachen University, Aachen, Germany
\\
\bigskip
* j.senk@fz-juelich.de

\end{flushleft}

\section*{Abstract}

Spatiotemporal patterns such as traveling waves are frequently observed
in recordings of neural activity. The mechanisms underlying the generation
of such patterns are largely unknown. Previous studies have investigated
the existence and uniqueness of different types of waves or bumps
of activity using neural-field models, phenomenological coarse-grained
descriptions of neural-network dynamics. But it remains unclear how
these insights can be transferred to more biologically realistic networks
of spiking neurons, where individual neurons fire irregularly. Here,
we employ mean-field theory to reduce a microscopic model of leaky
integrate-and-fire (LIF) neurons with distance-dependent connectivity
to an effective neural-field model. In contrast to existing phenomenological
descriptions, the dynamics in this neural-field model depends on the
mean and the variance in the synaptic input, both determining the
amplitude and the temporal structure of the resulting effective coupling
kernel. For the neural-field model we employ liner stability analysis
to derive conditions for the existence of spatial and temporal oscillations
and \textcolor{black}{wave trains, that is, temporally and spatially
periodic traveling waves}. We first prove that wave trains cannot
occur in a single homogeneous population of neurons, irrespective
of the form of distance dependence of the connection probability.
Compatible with the architecture of cortical neural networks, wave
trains emerge in two-population networks of excitatory and inhibitory
neurons as a combination of delay-induced temporal oscillations and
spatial oscillations due to distance-dependent connectivity profiles.
Finally, we demonstrate quantitative agreement between predictions
of the analytically tractable neural-field model and numerical simulations
of both networks of nonlinear rate-based units and networks of LIF
neurons.

\section{Introduction\label{sec:Introduction}}

Experimental recordings of neural activity frequently reveal spatiotemporal
patterns such as traveling waves propagating across the cortical surface
\cite{Rubino-2006_1549,Nauhaus09_70,Muller12_222,Sato12_218,Muller14_4675,Townsend15_4657,Zanos15_615,Denker18_1}
or within other brain regions such as the thalamus \cite{Kim95_1301,Muller12_222}
or the hippocampus \cite{Lubenov09_534}. These large-scale dynamical
phenomena are detected in local-field potentials (LFP) \cite{Riehle13_48}
and in the spiking activity \cite{Takahashi15} recorded with multi-electrode
arrays, by voltage-sensitive dye imaging \cite{Ferezou06_617}, or
by two-photon imaging monitoring the intracellular calcium concentration
\cite{Garaschuk00_452}. They have been reported in in-vitro and in
in-vivo experiments, in both anesthetized and awake states, and during
spontaneous as well as stimulus-evoked activity \cite{Muller12_222}.

\begin{figure}[t]
\begin{centering}
\includegraphics[width=0.85\textwidth]{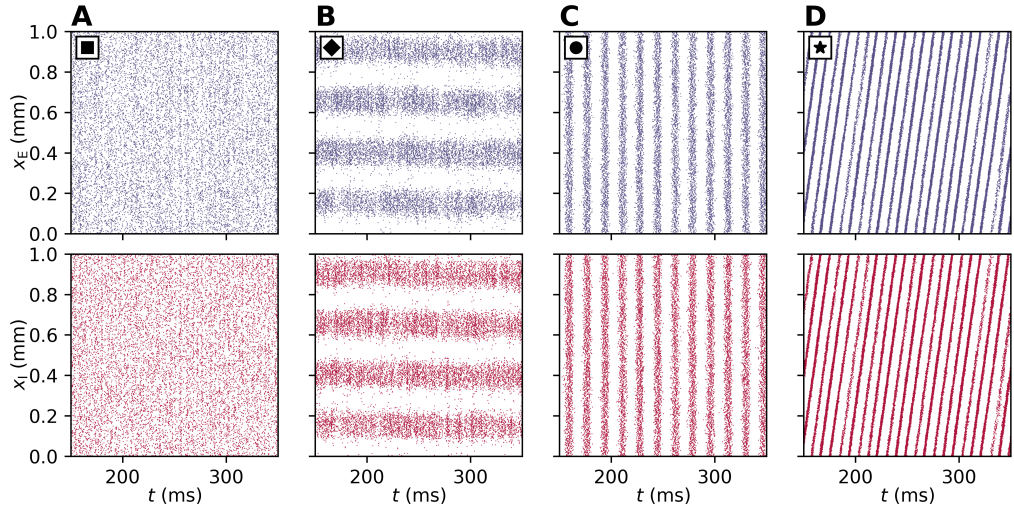}
\par\end{centering}
\begin{centering}
\par\end{centering}
\caption{\textbf{Spatiotemporal patterns in a spiking neural network model.}
Spiking activity of recurrently connected populations of excitatory
($\mathrm{E}$, blue) and inhibitory ($\mathrm{I}$, red) leaky integrate-and-fire
neurons. Each dot represents the spike-emission time of a particular
neuron. Neurons are positioned on a ring with a perimeter of $\unit[1]{mm}$.
Each neuron receives a fixed number of incoming connections from its
excitatory (inhibitory) neighbors uniformly and randomly drawn within
a distance of $R_{\mathrm{E}}$ ($R_{\mathrm{I}}$). The spike-transmission
delay $d$, the widths $R_{\mathrm{E}}$ and $R_{\mathrm{I}}$ of
the spatial connectivity profiles, and the relative inhibitory synaptic
weight $g$\textbf{ }are varied.\textbf{ A}~Asynchronous-irregular
activity ($d=\unit[1]{ms}$, $R_{\mathrm{E}}=R_{\mathrm{I}}=\unit[0.4]{mm}$,
$g=6$). \textbf{B~}Oscillations in space ($d=\unit[3]{ms}$, $R_{\mathrm{E}}=\unit[0.1]{mm}$,
$R_{\mathrm{I}}=\unit[0.15]{mm}$, $g=5$). \textbf{C}~Oscillations
in time ($d=\unit[6]{ms}$, $R_{\mathrm{E}}=R_{\mathrm{I}}=\unit[0.4]{mm}$,
$g=7$). \textbf{D}~Propagating waves ($d=\unit[3]{ms}$, $R_{\mathrm{E}}=\unit[0.2]{mm}$,
$R_{\mathrm{I}}=\unit[0.07]{mm}$, $g=5$). For remaining parameters,
see \prettyref{tab:parameters}. \label{fig:rasters_spatiotemporal_patterns}}
\end{figure}

Previous modeling studies have shown that networks of spiking neurons
with distance-dependent connectivity, extending in one- or two-dimensional
space, can exhibit a variety of such spatiotemporal patterns \cite{Mehring03_395,Yger2011_229,Voges12,Keane15_1591}.
For illustration, consider the example in \prettyref{fig:rasters_spatiotemporal_patterns}.
Depending on the choice of transmission delays, the spatial reach
of connections and the strength of inhibition, a network of leaky
integrate-and-fire (LIF) model neurons generates asynchronous-irregular
activity (A), spatial patterns that are persistent in time (B), spatially
uniform temporal oscillations (C), or propagating waves (D). Distance-dependent
connectivity is a prominent feature of biological networks. In the
neocortex, local connections are established within a radius of about
$\unit[500]{\mu m}$ around a neuron's cell body \cite{Voges10_277},
and the probability of two neurons being connected decays with distance
\cite{Hellwig00_111,Perin11,Schnepel15_3818}.

So far, the formation of spatiotemporal patterns in neural networks
has mainly been studied by means of phenomenological neural-field
models describing network dynamics at a macroscopic spatial scale
\cite{Wilson1972,Wilson1973,Amari77}. Such models can describe patterns
in recorded brain activity that are related to movement \cite{Erlhagen97}
or occur in response to a visual stimulus \cite{Bressloff15_e1004545}.
Neural-field models are formulated with continuous nonlinear integro-differential
equations for a spatially and temporally resolved activity variable
and usually possess an effective distance-dependent connectivity kernel.
These models provide insights into the existence and uniqueness of
diverse patterns which are stationary or nonstationary in space and
time, such as waves, wave fronts, bumps, pulses, and periodic patterns
(reviewed in \cite{Ermentrout98b,Coombes05,Wyller07_79,Coombes10_731,Bressloff12,Bressloff14_8,Coombes14}).
There are two main techniques for analyzing spatiotemporal patterns
in neural-field models \cite{Bressloff12}: First, in the constructive
approach introduced by Amari \cite{Amari77}, bump or wave solutions
are explicitly constructed by relating the spatial and temporal coordinates
of a nonlinear system (reviewed in \cite[Section 7]{Ermentrout98b}
and \cite[Sections 3-4]{Bressloff12}). Second, the emergence of periodic
patterns is studied with bifurcation theory as in the seminal works
of Ermentrout and Cowan \cite{Ermentrout79b,Ermentrout79a,Ermentrout80,Ermentrout80_323}.
In this latter framework, linear stability analysis is often employed
to detect pattern-forming instabilities and to derive conditions for
the onset of pattern formation (see for example \cite{Bressloff96_4644,Hutt03_351}
or the reviews \cite[Section 8]{Ermentrout98b} and \cite[Section 5]{Bressloff12}).
There are four general classes of states that can linearly bifurcate
from a homogeneous steady state: a new uniform stationary state, temporal
oscillations (spatially uniform and periodic in time, also known as
global `bulk oscillations' \cite{Bressloff08_41916}), spatial oscillations
(spatially periodic and stationary in time), and \textcolor{black}{wave
trains (spatially and temporally periodic; special type of traveling
waves)}, see \cite[Section 8]{Ermentrout98b} and \cite{Roxin05,Atay06_670,Venkov07_1}.
The analysis of these states is often called `(linear) Turing instability
analysis' \cite{Coombes05,Coombes07_51901,Venkov07_1} referring
to the work of Turing on patterns in reaction-diffusion systems \cite{Turing_52}.
The respective instabilities leading to these states are termed: a
firing rate instability, Hopf instability \cite{Kuramoto84_3}, Turing
instability, and Turing-Hopf \cite{Roxin05} or `wave' \cite{Hutt03_351}
instability. The instabilities generating temporally periodic patterns
(Hopf and Turing-Hopf instabilities) are known as `dynamic' \cite{Venkov07_1}
or `nonstationary' \cite{Hutt05_30} instabilities, in contrast
to `static' \cite{Venkov07_1} or `stationary' \cite{Hutt05_30}
instabilities generating temporally stationary patterns.

The emergence of pattern-forming instabilities has been investigated
with respect to system parameters such as the spatial reach of excitation
and inhibition in an effective connectivity profile \cite{Ermentrout98b};
specifically without transmission delays \cite{Wyller07_75,Folias12_895},
or with constant \cite{Roxin05,Roxin06}, distance-dependent \cite{Jirsa00_8462,Hutt03_351,Atay05_644,Atay06_670,Coombes07_51901,Bressloff08_41916,Hutt08_541,Bojak10_e1000653,Hutt10_55701}
or both types \cite{Veltz11_749,Veltz13_1566} of delays. \textcolor{black}{Faye
and Faugeras\cite{Faye10_561} show existence and uniqueness of solutions
and provide conditions for asymptotic stability of the trivial homogeneous
steady state of the corresponding linearized system using the Lyapunov
functional. The principle of linearized stability for such models
was proven by Veltz and Faugeras \cite{Veltz2011}. Dijkstra et al.
\cite{Dijkstra2015} provide a rigorous analysis of a one-dimensional
neural-field model revealing pitchfork-Hopf bifurcations. The existence
of standing waves emerging from a Turing bifurcation of the trivial
homogeneous steady state, in a linearized neural field model with
space-dependent delays on a sphere, was shown by Visser et al. \cite{Visser2017}.}

Neural-field models treat neural tissue as a continuous excitable
medium and describe neural activity in terms of a space and time dependent
real-valued quantity. Throughout the current work the spatial coordinate
refers to physical space, although in general it could also be interpreted
as feature space. At the microscopic scale, in contrast, neural
networks are composed of discrete units (neurons) – which interact
via occasional short stereotypical pulses (spikes) rather than continuous
quantities like firing rates. In the neocortex, spiking activity is
typically highly irregular and sparse \cite{Softky93,Brunel99}, with
weak pairwise correlations \cite{Ecker10}. To date, a rigorous link
between this microscopic level and the macroscopic description by
neural-field models is lacking \cite{Coombes10_731,Bressloff14_8,Hutt15_141,Montbrio15}.
While randomly connected spiking networks have been extensively analyzed
using mean-field approaches \cite{Amit-1997_373,Brunel99,Brunel00_183,Lindner05_061919,Delarue2015},
the theoretical understanding of spatially structured spiking networks
is still deficient. \textcolor{black}{A recent work in this direction
is Esnaola-Acebes et al. \cite{Esnaola-Acebes2017}, who investigate
ring networks of quadratic integrate-and-fire model neurons and provide
bifurcation diagrams showing temporal oscillations and bump states,
supported by both mathematical analysis and simulation.} But in general
it remains unclear how to qualitatively transfer insights on the formation
of spatiotemporal patterns from neural fields to networks of spiking
neurons. Moreover, it is unknown how the multitude of neuron, synapse
and connectivity parameters of spiking neural networks relates to
the effective parameters in neural-field models. A quantitative link
between the two levels of description is, for example, required for
adjusting parameters in a network of spiking neurons such that it
generates a specific type of spatiotemporal pattern, and to enable
model validation by comparison with experimental data.

Different efforts have already been undertaken to match spiking and
time-continuous rate models with spatial structure. Certain assumptions
and approximations allow the application of techniques for analyzing
spatiotemporal patterns developed for neural-field models. The above
mentioned constructive approach \cite{Amari77}, for example, can
be applied to networks of spiking neurons under the assumption that
every neuron spikes at most once, thus ignoring the sustained spike
generation and after-spike dynamics of biological neurons \cite{Golomb01_4179,Cremers02_1651,Osan02}.
A related simplification substitutes a spike train by an ansatz for
a wave front. This leads to a mean-field description of single-spike
activity often applied to a spike-response model \cite{Fohlmeister95_905,Kistler98_273,Kistler00_8834,Bressloff00_169}.
Traveling-wave solutions have also been proposed for a network of
coupled oscillators and a corresponding continuum model \cite{Crook97_161}.
In the framework of bifurcation theory, Roxin et al. \cite{Roxin05,Roxin06}
demonstrate a qualitative agreement between a neural-field model and
a numerically simulated network of Hodgkin-Huxley-type neurons in
terms of emerging spatiotemporal patterns. However, the authors do
not observe stable traveling waves in the spiking network, even though
the neural-field model predicts their occurrence. In the limit of
slow synaptic interactions, spiking dynamics can be reduced to a mean-firing-rate
model for studying bifurcations \cite{Ermentrout94_679,Bressloff98_2384,Bressloff00_820}.
An example is the lighthouse model \cite{Haken00_187,Haken00_96},
defined as a hybrid between a phase oscillator and a firing-rate model,
that reduces to a pure rate model for slow synapses \cite{Chow06_552}.
Laing and Chow \cite{Laing01} demonstrate a bump solution in a spiking
network and discuss a corresponding rate model. Recently, the group
around Doiron and Rosenbaum explored in a sequence of studies spatially
structured networks of LIF neurons without transmission delays in
the continuum limit with respect to the spatial widths of connectivity.
The authors focus on the existence of the balanced state \cite{Rosenbaum14},
the structure of correlations in the spiking activity \cite{Rosenbaum16_107},
and bifurcations in the linearized dynamics in relation to network
computations \cite{Pyle17_18103}. \textcolor{black}{Spreizer et
al. \cite{Spreizer18_bioarxiv_v1} further demonstrate that spatiotemporal
activity sequences can be induced by anisotropic but spatially correlated
connectivity.} Kriener et al. \cite{Kriener14} employ static mean-field
theory and extend the linearization of a network of LIF neurons with
constant delays as described by Brunel \cite{Brunel00_183}, to spatially
structured networks. The work derives conditions for the appearance
of spontaneous symmetry breaking that leads to stationary periodic
bump solutions (spatial oscillations), and distinguish between the
mean-driven and the fluctuation driven regime. \textcolor{black}{A
coarse-graining procedure for a ring network of modified binary neurons
with refractoriness was presented by Avitable and Wedgwood \cite{Avitable2017}.
By combining analytical and numerical analysis they show existence
of bumps and traveling waves.}

Despite these previous works on spatially structured network models
of spiking neurons and attempts to link them with neural-field models,
there still exists no systematic way of mapping parameters between
these models. Furthermore, none of these studies focuses on uncovering
the underlying mechanism of wave trains in spiking networks. In the
present work we establish the so far missing, quantitative link between
a sparsely connected network of spiking LIF neurons with spatial structure
and a typical neural-field model. An explicit parameter mapping between
the two levels of description allows us to study the origin of spatiotemporal
patterns analytically in the neural-field model using linear stability
analysis, and to reproduce the predicted patterns in spiking activity.
We employ mean-field theory to derive the neural-field model as an
effective rate model depending on the dynamical working point of the
network that is characterized by both the mean and the variance of
the synaptic input. The rate model accounts for biological constraints
such as a static weight that is either positive (excitatory) or negative
(inhibitory) and a spatial profile that can be interpreted as a distance-dependent
connection probability. Given these constraints, we show that wave
trains cannot occur in a single homogeneous population irrespective
of the shape of distance-dependent connection probability. For two-population
networks of excitatory and inhibitory neurons, in contrast, wave trains
emerge for specific types of spatial profiles and for sufficiently
large delays, as shown in \prettyref{fig:rasters_spatiotemporal_patterns}D.

The remainder of the study is structured as follows: In \nameref{sec:Results}
we derive the conditions for the existence of wave trains for a typical
neural-field model by linear stability analysis, present an effective
model corresponding to the microscopic description of spiking neurons,
compare the two models, and show simulation results for validation.
In \nameref{sec:Discussion} we put our results in the context of
previous literature. Finally, \nameref{sec:Methods} contains details
on our approach. An account of the presented work has previously been
published in abstract form in \cite{Senk17_cns_P94}.

\section{Results\label{sec:Results}}

\begin{figure}[t]
\begin{centering}
\includegraphics[width=0.55\textwidth]{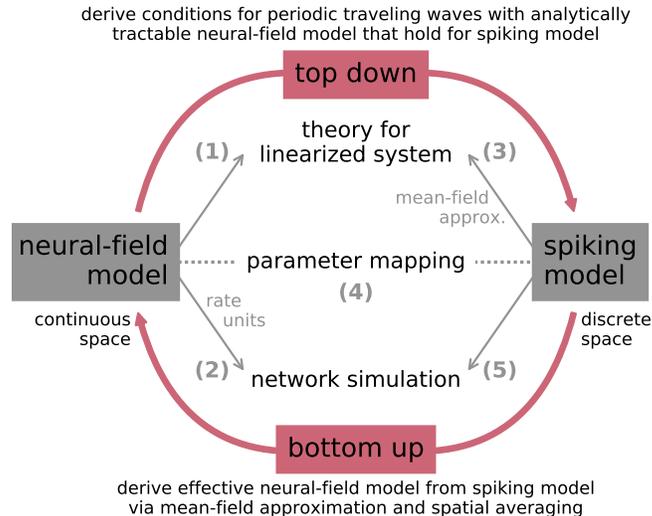}
\par\end{centering}
\centering{}\caption{\textbf{Mapping microscopic single-neuron dynamics to spatially averaged
population dynamics. (1)}~Conditions for wave trains in a neural-field
model. \textbf{(2)}~Network simulation of discrete nonlinear rate
neurons. \textbf{(3)}~Mean-field approximation of the spiking model
and spatial averaging lead to an effective linearized continuous system.
\textbf{(4)}~Parameter mapping between spiking and neural-field model.
\textbf{(5)}~Network simulation of spiking neurons and validation
of analytical results.\label{fig:overview}}
\end{figure}

We aim to establish a mapping between two different levels of description
for spatially structured neural systems to which we refer as ‘neural-field
model’ and ‘spiking model’ based on the initial model assumptions.
While the neural-field model describes neural activity as a quantity
that is continuous in space and time, the spiking model assumes a
network of recurrently connected spiking model neurons in discrete
space. Our methodological approach for mapping between these two models,
as well as the structure of this section, are illustrated in \prettyref{fig:overview}.
(1) We start in Sections \ref{sub:Linear-stability-analysis}-\ref{sub:Application-to-a-concrete-network}
with linear stability analysis of a typical neural-field model that
is a well-known and analytically tractable rate equation. This approach
builds on existing literature (cf. \cite[Section 8]{Ermentrout98b}
and \cite[Section 5]{Bressloff12}) and introduces the concepts of
our study with modest mathematical efforts. We analyze the neural-field
model for one and two populations and derive conditions for the occurrence
of wave trains based on spatial connectivity profiles and transmission
delays. (2) In \prettyref{sub:Network-simulation-with-nonlinear-rate}
we continue with simulations of a discrete version of the neural-field
model, a network of nonlinear rate-based units, and show that the
results from our linear analysis indeed accurately predict transitions
between network states (homogeneously steady, spatial oscillations,
temporal oscillations, waves). (3) Then, in \prettyref{sub:Linearization-of-spiking}
we linearize the population dynamics of networks of discrete spiking
leaky integrate-and-fire (LIF) neurons using mean-field theory and
derive expressions similar to the neural-field model. (4) Thus, both
the linearized neural-field and spiking models can be treated in a
conceptually similar manner, with the exception of an effective coupling
kernel which is mathematically more involved for the spiking model.
In \prettyref{sub:Comparison-of-models} we perform a parameter mapping
between the biophysically motivated parameters of the spiking model
and the effective parameters of a neural-field model. (5) Finally,
in \prettyref{sub:Validation-in-simulation} we demonstrate that the
insights obtained in the analysis of the neural-field model apply
to networks of simulated LIF neurons: The bifurcations indeed appear
at the theoretically predicted parameter values.

In summary, the mapping of a microscopic spiking network model to
a continuum neural-field model (bottom up) allows us to transfer analytically
derived insights from the neural-field model directly to the spiking
model (top down).

\subsection{Linear stability analysis of a neural-field model\label{sub:Linear-stability-analysis}}

\begin{figure}[t]
\begin{centering}
\includegraphics[width=0.85\textwidth]{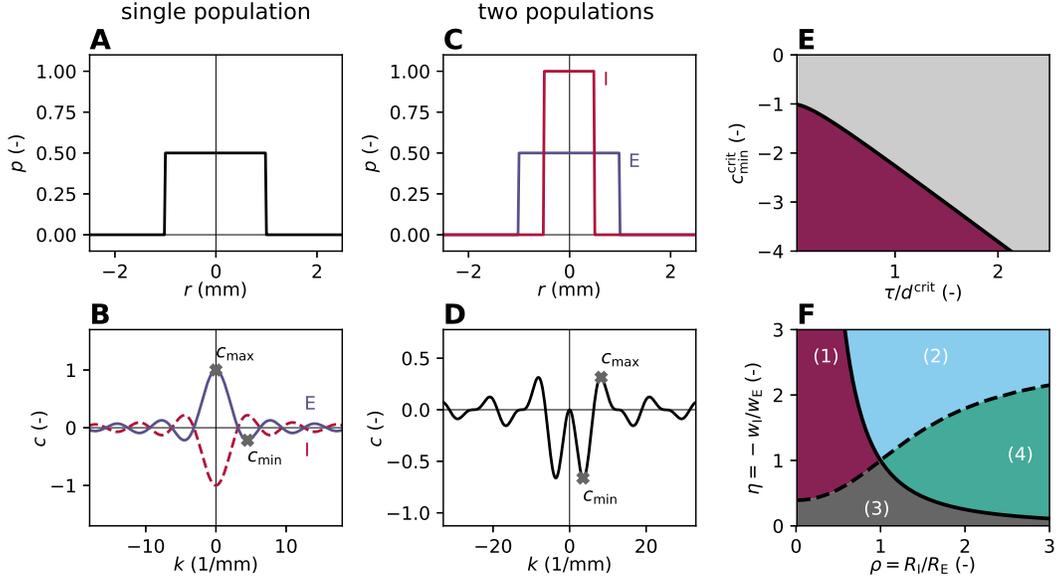}
\par\end{centering}

\centering{}\caption{\textbf{Effective profile yields conditions for wave trains.} \textbf{A}~Boxcar-shaped
spatial profile $p$ of width $R=\unit[1]{mm}$ for a single population.
\textbf{B}~Effective profile $c$ (blue curve) denotes Fourier transform
of spatial profile $\widehat{p}$ times positive weight $w_{\mathrm{E}}=1$.
Gray crosses indicate maximum $\mathrm{c_{\mathrm{max}}}$ and minimum
$c_{\mathrm{min}}$. Same spatial profile but with negative weight
($w_{\mathrm{I}}=-w_{\mathrm{E}}$) yields mirrored curve (red, dashed
line). \textbf{C~}Spatial profiles of different widths for two populations
$\mathrm{E}$ ($R_{\mathrm{E}}=\unit[1]{mm}$, blue) and $\mathrm{I}$
($R_{\mathrm{I}}=\unit[0.5]{mm}$, red). \textbf{D}~Effective profile:
$c\left(k\right)=w_{\mathrm{E}}\widehat{p}_{\mathrm{E}}\left(k\right)+w_{\mathrm{I}}\widehat{p}_{\mathrm{I}}\left(k\right)$.
\textbf{E}~Transition curve $c_{\mathrm{min}}^{\mathrm{crit}}(\tau/d^{\mathrm{crit}})$
given by \prettyref{eq:P_crit} for Hopf bifurcation indicating onset
of delay-induced oscillations (appearing in purple region) with time
constant $\tau$ and delay $d$. \textbf{F}~Transition curves for
relative width $\rho=R_{\mathrm{I}}/R_{\mathrm{E}}$ and relative
weight $\eta=-w_{\mathrm{I}}/w_{\mathrm{E}}$. Colored regions indicate
which extremum, the minimum $c_{\mathrm{min}}$ or the maximum $c_{\mathrm{max}}$,
has larger absolute value and if the dominant one occurs at $k=0$
or at $k>0$. Purple (1): $c_{\mathrm{min}}$ appears at $k_{\mathrm{min}}>0$.
Light blue (2): $c_{\mathrm{min}}$ appears at $k_{\mathrm{min}}=0$.
Dark gray (3): $c_{\mathrm{max}}$ appears at $k_{\mathrm{max}}=0$.
Green (4) $c_{\mathrm{max}}$ appears at $k_{\mathrm{max}}>0$. \label{fig:conditions_pTW}}
\end{figure}

We first consider a neural-field model with a single population defined
as a continuous excitable medium with a translation-invariant interaction
kernel and delayed interaction in one spatial dimension. The dynamics
follows an integro-differential equation

\begin{equation}
\tau\,\frac{\partial u}{\mathrm{\partial}t}\left(x,t\right)+u\left(x,t\right)=\int_{-\infty}^{\infty}m\left(x-y\right)\,\psi\left(u\left(y,t-d\right)\right)\,\mathrm{d}y.\label{eq:intdiff}
\end{equation}
The variable $u$ describes the activity of the neural population
at position $x$ at time $t$. Here $\tau>0$ denotes a time constant
and $d>0$ a transmission delay. The function $\psi$ describes the
nonlinear transformation of the output activity $u$ if considered
as input to the neural field. The function $m$ specifies the translation-invariant
connectivity depending only on the displacement $r=x-y$ where $x$
and $y$ denote neuron positions. Earlier studies show that specific
choices for connectivities $P$ and nonlinear transformations $\psi$
result in spatiotemporal patterns such as waves or bumps \cite{Ermentrout98b,Coombes05,Wyller07_79,Coombes10_731,Bressloff12,Bressloff14_8,Coombes14}.

Here, we assume that the connectivity $m$ is isotropic and define
$m\left(r\right)\coloneqq w\,p\left(r\right)$. The scalar weight
$w$ can either be positive (excitatory) or negative (inhibitory).
The spatial profile $p(r)$ is a symmetric probability density function
with the properties $p\left(r\right)=p\left(-r\right)$, $p(r)>0$
for $r\in\left(-\infty,\infty\right)$ and $\int_{-\infty}^{\infty}p\left(r\right)\,\mathrm{d}r=1$.
\prettyref{fig:conditions_pTW}A shows, as an example, a boxcar-shaped
spatial profile with width $R$, defined by $p\left(r\right)=\frac{1}{2R}\Theta\left(R-\left|r\right|\right)$
where $\Theta$ denotes the Heaviside function.

Throughout this study we investigate bifurcations of the system \prettyref{eq:intdiff}
between a state of spatially and temporally homogeneous activity $u(x,t)=u_{0}$
to states where the activity shows structure in the temporal domain,
in the spatial domain, or both. For this purpose we use Turing instability
analysis \cite{Bressloff96_4644,Hutt03_351,Coombes05}. Initially
we assume that the model parameters are chosen such that the homogeneous
solution is locally asymptotically stable, implying that small perturbations
away from $u_{0}$ will relax back to this baseline. We ask the question:
In which regions of the parameter space ($R$, $d$, $w$, $\psi$)
is the stability of the homogeneous solution lost? To this end we
linearize around the steady state and denote deviations $\delta u(t)=u(t)-u_{0}$.
Without loss of generality we assume the slope $\psi^{\prime}(u_{0})$
of the gain function to be unity; a non-zero slope can be absorbed
into a redefinition of $w$. \textcolor{black}{We here use a gain function
that allows $u$ to become negative. Likewise, one can treat nonlinear
gain functions $\psi$ that are strictly positive (see, e.g., \cite{Roxin06}).
These two conventions can be mapped to one another by a suitable shift
of variables. In either case, after linearization the deviation $\delta u$
does not have a definite sign. Because the resulting linear system
is invariant with respect to translations in time and space, its eigenmodes
are Fourier-Laplace modes of the form}

\begin{align}
\delta u\left(x,t\right) & =\mathrm{e}^{\mathrm{i}kx}\mathrm{e}^{\lambda t},\label{eq:Fourier_mode}
\end{align}
where the wave number $k\in\mathbb{R}$ is real and the temporal eigenvalue
$\lambda\in\mathbb{C}$ is complex. Solutions constructed from these
eigenmodes can oscillate in time and space, and exponentially grow
or decay in time. The characteristic equation (see \prettyref{eq:derivation_char_eq}
in \nameref{sec:Methods})

\begin{equation}
\left(1+\tau\lambda\right)\,\mathrm{e}^{\lambda d}=c\left(k\right),\label{eq:char_eq}
\end{equation}
comprises the effective profile $c\left(k\right)\coloneqq\widehat{m}\left(k\right)\coloneqq w\widehat{p}\left(k\right)$.
The Fourier transform of the spatial profile is denoted by $\widehat{p}\left(k\right)$
which, by its definition as a probability density, is maximal at $k=0$
with $\widehat{p}\left(0\right)=1$ (see Eqs \ref{eq:abs_p_hat1}
and \ref{eq:abs_p_hat2} in \nameref{sec:Methods}). The effective
profile for the boxcar-shaped spatial profile is shown in \prettyref{fig:conditions_pTW}B,
for excitatory and inhibitory weights with absolute magnitudes of
unity.

We next extend the system to two populations, an excitatory one denoted
by $\mathrm{E}$, and an inhibitory one by $\mathrm{I}$. Time constants
$\tau$ and delays $d$ are assumed to be equal for both populations,
but $u$ becomes a vector, $u=\left(u_{\mathrm{E}},u_{\mathrm{I}}\right)^{T}$,
and the connectivity $m\left(r\right)$ a matrix

\begin{equation}
M\left(r\right)=\left(\begin{array}{cc}
w_{\mathrm{EE}}\,p_{\mathrm{EE}}\left(r\right) & w_{\mathrm{EI}}\,p_{\mathrm{EI}}\left(r\right)\\
w_{\mathrm{IE}}\,p_{\mathrm{IE}}\left(r\right) & w_{\mathrm{II}}\,p_{\mathrm{II}}\left(r\right)
\end{array}\right).\label{eq:connectivity_rate}
\end{equation}
The linearized system again possesses the same symmetries as the counterpart
for a single population so that the eigenmodes for the deviation from
the stationary state are of the form $\delta u\left(x,t\right)=ve^{\mathrm{i}kx}\mathrm{e^{\lambda t}}$
with $v$ denoting a constant vector. Hence, we arrive at an auxiliary
eigenvalue problem (see \prettyref{eq:aux_eig_prob} in \nameref{sec:Methods})
with the two eigenvalues

\begin{equation}
c_{1,2}\left(k\right)=\frac{1}{2}\left(w_{\mathrm{EE}}\,\widehat{p}_{\mathrm{EE}}\left(k\right)+w_{\mathrm{II}}\,\widehat{p}{}_{\mathrm{II}}\left(k\right)\pm\sqrt{D}\right),\label{eq:eigenvalues_P_hat}
\end{equation}
where

\begin{equation}
D=\left(w_{\mathrm{EE}}\,\widehat{p}_{\mathrm{EE}}\left(k\right)-w_{\mathrm{II}}\,\widehat{p}_{\mathrm{II}}\left(k\right)\right)^{2}+4w_{\mathrm{EI}}\,\widehat{p}_{\mathrm{EI}}\left(k\right)\,w_{\mathrm{IE}}\,\widehat{p}_{\mathrm{IE}}\left(k\right).
\end{equation}
These two eigenvalues play the same role as the effective profile
$c$ in the one-population case above. As a consequence, the same
characteristic equation \prettyref{eq:char_eq} holds for both the
one- and the two-population system, \textcolor{black}{as a scalar and
two-dimensional vector equation, respectively.}

In the following example we restrict the weights and the spatial profiles
to be uniquely determined by the source population alone, denoted
by $w_{\mathrm{\alpha E}}=:w_{\mathrm{E}}$, $w_{\mathrm{\alpha I}}=:w_{\mathrm{I}}$
for $\alpha\in\{\mathrm{E},\mathrm{I}\}$. An illustration of the
two spatial profiles of different widths $R_{\mathrm{E}}$ and $R_{\mathrm{I}}$
is shown in \prettyref{fig:conditions_pTW}C. \textcolor{brown}{}\textcolor{black}{The
respective effective profile \prettyref{eq:eigenvalues_P_hat} reduces
to $c_{1}\left(k\right)=w_{\mathrm{E}}\widehat{p}_{\mathrm{E}}\left(k\right)+w_{\mathrm{I}}\widehat{p}_{\mathrm{I}}\left(k\right)=:c\left(k\right)$,
and is shown in \prettyref{fig:conditions_pTW}D; $c_{2}\equiv0$
for all $k.$}\textcolor{brown}{}

The characteristic equation \prettyref{eq:char_eq} can be solved
for the eigenvalues $\lambda$ by using the Lambert W function defined
as $z=W\left(z\right)\mathrm{e}^{W\left(z\right)}$ for $z\in\mathbb{C}$
\cite{Corless96_329}. The Lambert W function has infinitely many
branches, indexed by $b$, and the branch with the largest real part
is denoted the principle branch ($b=0$), see Eqs \ref{eq:largest_real_part_1}-\ref{eq:largest_real_part_2}
in \nameref{sec:Methods} for a proof. The characteristic equation
determines the temporal eigenvalues (see \prettyref{eq:char_eq_to_LW}
in \nameref{sec:Methods} and compare with \cite{Veltz13_1566})

\begin{equation}
\lambda_{b}(k)=-\frac{1}{\tau}+\frac{1}{d}\,W_{b}\left(c\left(k\right)\frac{d}{\tau}\mathrm{e}^{\frac{d}{\tau}}\right).\label{eq:eigenvalues_rate_LW}
\end{equation}

\textcolor{black}{As shown by Veltz and Faugeras \cite{Veltz2011},
linearized stability of the homogeneous steady state of \prettyref{eq:intdiff}
is fully determined by the eigenvalues \prettyref{eq:eigenvalues_rate_LW}.
These authors assume an open bounded domain and provide an example
of a one-dimensional ring network. In our theoretical analysis, we
decide to work with the infinite domain for technical convenience.
Formulating the problem on a ring with periodic boundary conditions
would not change any of our conclusions. The added value of their
approach is the possibility to justify all steps in a mathematically
rigorous way. In our approach, the temporal eigenvalue $\lambda_{b}$
is a continuous function of the wave number $k.$ On a bounded domain
with periodic boundary conditions, one obtains a discrete set of wave
numbers $k$. Since the temporal eigenvalue $\lambda_{b}$ varies
on a much slower scale compared to the resolution of discrete wave
numbers $k$, this change does not have any qualitative effect on
the resulting dynamics. The infinite domain, however, allows us to
easily incorporate spatial profiles with unbounded support. For profiles
with unbounded support that decay to zero fast enough, the theoretical
prediction of the frequency of oscillation can be regarded as an approximation
of the dynamics on a ring with periodic boundary conditions.}

\subsection{Conditions for spatial and temporal oscillations, and wave trains\label{sub:Conditions-for-traveling}}

 The homogeneous (steady) state of our system is locally asymptotically
stable if the real parts of all eigenvalues $\lambda_{b}$ are negative
\begin{equation}
\operatorname{Re}\left[W_{b}\left(c\left(k\right)\frac{d}{\tau}\mathrm{e}^{\frac{d}{\tau}}\right)\right]<\frac{d}{\tau},{\normalcolor }\label{eq:stability_condition}
\end{equation}
for all branches $b$ of the Lambert W function. The system loses
stability when the real part of the eigenvalue $\lambda_{0}$ on the
principle branch becomes positive at a certain $k=k^{*}$. Such instabilities
may occur either for a positive or a negative argument of the Lambert
W function.

We denote the maximum of $c$ as $c_{\mathrm{max}}$ and the minimum
as $\mathrm{c_{\mathrm{min}}}$ occurring at $k_{\mathrm{max}}$ and
$k\mathrm{_{min}}$, respectively, as indicated in \prettyref{fig:conditions_pTW}B
and D. \textcolor{black}{The system becomes unstable for a positive
argument of $W$ if $c_{\mathrm{max}}=1$ where $W\left(\frac{d}{\tau}\mathrm{e}^{\frac{d}{\tau}}\right)=\frac{d}{\tau}$
by the definition of the Lambert W function; so equality holds in
\prettyref{eq:stability_condition} independent of the values $d$
and $\tau$. The imaginary part of $\lambda_{0}$ is zero at such
a transition because the principal branch of the Lambert W function
has real values for positive real arguments.} If the instability
appears at a wave number $k^{\ast}=0$, the population activity is
collectively destabilized. This transition corresponds in networks
of binary neurons and of spiking neurons to the transition between
the asynchronous irregular (AI) state and the synchronous regular
(SR) state, where the system ceases to be stabilized by negative feedback
and leaves the balanced state \cite{Vreeswijk96,Brunel00_183}.
If this transition appears at a wave number $k^{*}>0$, it follows
from \prettyref{eq:Fourier_mode} that the activity shows spatial
oscillations that grow exponentially in time.

For a negative argument of $W$ of less than $-1/e$, the eigenvalues
\prettyref{eq:eigenvalues_rate_LW} come in complex conjugate pairs.
The real part of $\lambda_{0}$ becomes positive if the condition
\begin{equation}
\operatorname{Re}\left[W_{0}\left(c_{\mathrm{min}}\frac{d}{\tau}\mathrm{e}^{\frac{d}{\tau}}\right)\right]=\frac{d}{\tau}
\end{equation}
is fulfilled with a negative $c_{\mathrm{min}}<-1$. Because the eigenvalues
have non-zero imaginary parts, this transition corresponds to a Hopf
bifurcation and the onset of temporal oscillations. The condition
for this bifurcation has been derived earlier \cite[Eq 10]{Helias13_023002}

\begin{equation}
\frac{d^{\mathrm{crit}}}{\tau}=\frac{\pi-\arctan\left(\sqrt{c_{\mathrm{min}}^{\mathrm{crit^{2}}}-1}\right)}{\sqrt{c_{\mathrm{min}}^{\mathrm{crit}^{2}}-1}}.\label{eq:P_crit}
\end{equation}
Here, $d^{\mathrm{crit}}$ denotes the critical delay and $c_{\mathrm{min}}^{\mathrm{crit}}$
a critical minimum of the effective profile for points on the transition
curve. The system is stable for $c_{\mathrm{min}}>-1$ for all delays.
For larger absolute values of $c_{\mathrm{min}}$, the bifurcation
point is given by the critical value of the ratio between the time
constant and the delay, shown in \prettyref{fig:conditions_pTW}E.
If the transition occurs at $k^{\ast}=0$, temporal oscillations emerge
in which all neurons of the population oscillate in phase (`bulk
oscillations' \cite{Bressloff08_41916}). In spiking networks this
Hopf bifurcation corresponds to the transition from the AI regime
to the state termed ‘synchronous irregular fast (SI fast)’ \cite{Brunel99}.
If the transition appears for $k^{*}>0$, spatial and temporal oscillations
occur simultaneously. This phenomenon is known as `wave trains',
see \cite[Section 8]{Ermentrout98b} and \cite{Roxin05,Atay06_670,Venkov07_1}.
For the case that the system becomes unstable due to $c_{\mathrm{max}}$
reaching unity, the transition curve in \prettyref{fig:conditions_pTW}E
also provides a lower bound $c_{\mathrm{min}}^{\mathrm{crit}}(\tau/d^{\mathrm{crit}})$
above which temporal oscillations do not occur prior to the transition
due to $c_{\mathrm{max}}$.

\textcolor{brown}{}\textcolor{black}{A Hopf bifurcation can give rise
to either an asymptotically stable or unstable limit cycle, in the
super- or subcritical case, respectively. In our analysis we only
identify the Hopf bifurcation point by checking when a complex conjugate
pair of eigenvalues crosses the imaginary axis, and therefore cannot
predict the stability of the emerging limit cycle. If we, however,
in the simulation observe the transition from an asymptotically stable
homogeneous steady state, corresponding to the asynchronous irregular
regime, to spatiotemporal patterns, corresponding to a stable limit
cycle, and make sure that the initial conditions are close enough
to the homogeneous steady state, we know that the bifurcation we see
is indeed a supercritical Hopf bifurcation. The analytical conditions
for wave trains that we derive are necessary, but not sufficient.}\textcolor{brown}{}

\begin{table}[t]
\begin{centering}
\begin{tabular}{|c|c|c|c|c|}
\hline 
 & homogeneous & spatial oscillations & temporal oscillations & wave trains\tabularnewline
\hline 
\mytabspace{$c_{\mathrm{max}}$} & $<1$ & \cellcolor{lightgray}$1$ & $<1$ & $<1$\tabularnewline
\hline 
\mytabspace{$c_{\mathrm{min}}$} & $>c_{\mathrm{min}}^{\mathrm{crit}}$ & $>c_{\mathrm{min}}^{\mathrm{crit}}$ & \cellcolor{lightgray}$c_{\mathrm{min}}^{\mathrm{crit}}$ & \cellcolor{lightgray}$c_{\mathrm{min}}^{\mathrm{crit}}$\tabularnewline
\hline 
\mytabspace{$d$} & $<d^{\mathrm{crit}}$ & $<d^{\mathrm{crit}}$ & \cellcolor{lightgray}$d^{\mathrm{crit}}$ & \cellcolor{lightgray}$d^{\mathrm{crit}}$\tabularnewline
\hline 
\mytabspace{$k^{*}$} & - & \cellcolor{lightgray}$>0$ & \cellcolor{lightgray}$0$ & \cellcolor{lightgray}$>0$\tabularnewline
\hline 
\end{tabular}
\par\end{centering}
\caption{\textbf{Conditions for the onset of spatial and temporal oscillations,
and wave trains. }Gray cells in each column indicate the conditions
required for the instability causing the bifurcation. White cells
denote the conditions for the respective other bifurcation not to
occur. Last row indicates whether the bifurcation happens for zero
or nonzero wave number $k^{*}$. Here $d^{\mathrm{crit}}$ and $c_{\mathrm{min}}^{\mathrm{crit}}$,
as defined in \prettyref{eq:P_crit} and shown in \prettyref{fig:conditions_pTW}E,
denote the critical delay and the minimum of the effective profile
on the transition curve for a Hopf bifurcation. \label{tab:conditions}}
\end{table}

In summary, the system is stable if $c_{\mathrm{max}}<1$ and $c_{\mathrm{min}}>c_{\mathrm{min}}^{\mathrm{crit}}(\tau/d^{\mathrm{crit}})$.
For transitions occurring at either $c_{\mathrm{max}}=1$ or $c_{\mathrm{min}}=c_{\mathrm{min}}^{\mathrm{crit}}(\tau/d^{\mathrm{crit}})$
we distinguish between solutions with $k^{*}=0$ or $k^{*}>0$. In
\prettyref{tab:conditions} we provide an overview of the conditions
for bifurcations leading to spatial, temporal, or spatiotemporal oscillatory
states. These conditions imply that a one-population neural-field
model does not permit wave trains, which follows from the fact that
the absolute value of $\widehat{p}$ is strictly maximal at $k=0$
(see Eqs \ref{eq:abs_p_hat1}–\ref{eq:abs_p_hat2} in \nameref{sec:Methods}).
For a purely excitatory population ($w>0$) the critical minimum $c_{\mathrm{min}}^{\mathrm{crit}}(\tau/d^{\mathrm{crit}})$
therefore cannot be reached while keeping the maximum $c_{\mathrm{max}}$
stable as $c_{\mathrm{max}}>\left|c_{\mathrm{min}}\right|$. For a
purely inhibitory population ($w<0$), the condition $k_{\mathrm{min}}>0$
is not fulfilled because $c_{\mathrm{min}}$ occurs at $k=0$ as $\widehat{p}$
has its global maximum at the origin.

For a neural-field model accounting for both excitation and inhibition,
however, we can select shapes and parameters of the spatial profiles,
weights and the delay that fulfill the conditions for the onset of
wave trains as demonstrated by example in the next section.

\subsection{Application to a network with excitatory and inhibitory populations\label{sub:Application-to-a-concrete-network}}

Based on the conditions derived in the previous section, the minimal
network in conformity with Dale's principle in which wave trains can
occur consists of one excitatory ($\mathrm{E}$) and one inhibitory
($\mathrm{I}$) population. As in the example in \prettyref{sub:Linear-stability-analysis},
we assume that the connection weights and widths of boxcar-shaped
spatial profiles only depend on the source population. The effective
profile \prettyref{eq:eigenvalues_P_hat} in this case is

\begin{equation}
c\left(k\right)=w_{\mathrm{E}}\,\frac{\sin\left(R_{\mathrm{E}}k\right)}{R_{\mathrm{E}}k}+w_{\mathrm{I}}\,\frac{\sin\left(R_{\mathrm{I}}k\right)}{R_{\mathrm{I}}k},\label{eq:P_hat_EI}
\end{equation}
and positive and negative peaks of the profile are responsible for
bifurcations to spatial or temporal oscillations or wave solutions,
respectively. The previous section derives that in particular the
position and height of the minima and maxima of the effective profile
are decisive. To assess parameter ranges in which the peaks of the
effective profile \prettyref{eq:P_hat_EI} change qualitatively, we
introduce the relative width $\rho\coloneqq R_{\mathrm{I}}/R_{\mathrm{E}}>0$
and the relative weight $\eta\coloneqq-w_{\mathrm{I}}/w_{\mathrm{E}}>0$,
divide $c\left(k\right)$ by $w_{\mathrm{E}}$ and introduce the rescaled
wave number $\kappa=R_{\mathrm{E}}k$ to arrive at the dimensionless
reduced profile

\begin{equation}
\tilde{c}\left(\kappa\right)=\frac{\sin\left(\kappa\right)}{\kappa}-\eta\frac{\sin\left(\rho\kappa\right)}{\rho\kappa},\label{eq:reduced_effective_profile}
\end{equation}
which simplifies the following analysis.

Our aim is to divide the parameter space $\left(\rho,\eta\right)$
into regions that have qualitatively similar shapes of the effective
profile. The \nameref{sec:Methods} section describes the derivation
of transition curves and \prettyref{fig:conditions_pTW}F illustrates
the resulting parameter space. Above the first transition curve $\eta_{t1}\left(\rho\right)$
(dashed curve, see \prettyref{eq:transition_line1} in \nameref{sec:Methods}),
the absolute value of $\widetilde{c}_{\mathrm{min}}$ is larger than
$\widetilde{c}_{\mathrm{max}}$ (regions 1 and 2), and vice versa
below this curve (regions 3 and 4). The second transition curve $\eta_{t2}\left(\rho\right)$
(solid curve, see \prettyref{eq:transition_line2} in \nameref{sec:Methods})
indicates whether the extremum with the largest absolute value occurs
at $k=0$ (regions 2 and 3) or at $k>0$ (regions 1 and 4). The diagram
provides the necessary conditions and corresponding parameter combinations
required for both spatial and spatiotemporal patterns, purely based
on the relative weights and the relative widths which determine the
effective profile. The analysis shows that wave trains require wider
excitation than inhibition, $\rho<1$, because only this relation
simultaneously realizes a minimum at a non-zero wave number $k^{\ast}$
and a maximum with a peak below unity (see \prettyref{tab:conditions}). 

A neural-field model exhibiting wave trains can therefore be constructed
at will by first selecting a point within region 1 of \prettyref{fig:conditions_pTW}F
where $\rho<1$ and $\eta$ ensures that $\left|\widetilde{c}_{\mathrm{min}}\right|>\widetilde{c}_{\mathrm{max}}$.
Next, $c$ is fixed by scaling $\widetilde{c}$ with the absolute
weight $w_{\mathrm{E}}$ such that $c_{\mathrm{max}}<1$ for a stable
bump solution and $c_{\mathrm{min}}<-1$ for a Hopf bifurcation. Finally
a delay $d>d^{\mathrm{crit}}$ specifies a point below the bifurcation
curve shown in \prettyref{fig:conditions_pTW}E, given by the sufficient
condition for the Hopf bifurcation in \prettyref{eq:P_crit}. Likewise,
solutions for purely temporal oscillations appear in region 2, where
$c_{\mathrm{min}}<-1$ is attained at a vanishing wave number $k$
and a delay $d>d^{\mathrm{crit}}$; in addition $c_{\mathrm{max}}<1$
ensures absence of the other bifurcation into spatial oscillations.
For purely spatial oscillations, however, the comparison of the absolute
values of $\widetilde{c}_{\mathrm{min}}$ and $\widetilde{c}_{\mathrm{max}}$
is not sufficient; it is hence not sufficient to rely on the dashed
curve separating regions 2 and 4 in \prettyref{fig:conditions_pTW}F.
A loss of stability due to $c_{\mathrm{max}}>1$ can emerge not only
in region 4 but also in region 2, because even if $\left|c_{\mathrm{min}}\right|>c_{\mathrm{max}}$,
stability of $c_{\mathrm{min}}$ can be ensured by a sufficiently
short delay $d<d^{\mathrm{crit}}$, as shown in \prettyref{tab:conditions}.

\subsection{Network simulation with nonlinear rate neurons\label{sub:Network-simulation-with-nonlinear-rate}}

\textcolor{black}{So far we have only considered a mathematical description
of the nonlinear system with time and space represented by continuous
variables and analytically analyzed its properties using linear stability
analysis.}\textcolor{black}{{} }Next, we test the derived conditions
for the onset of oscillations, summarized in \prettyref{tab:conditions},
for a nonlinear, discrete system in the continuum limit. We here
consider a network of $N_{\mathrm{E}}=4,000$ excitatory ($\mathrm{E}$)
and $N_{\mathrm{I}}=1,000$ inhibitory ($\mathrm{I}$) rate neurons
described by a discrete version of the neural-field equation \prettyref{eq:intdiff}
(see \prettyref{tab:model_description} for details). The model neurons
within each population are positioned on a ring of perimeter $L=\unit[1]{mm}$
as described in \prettyref{sec:Methods}. \textcolor{black}{We choose
periodic boundary conditions, i.e., the ring topology, due to the
inevitably finite size of the discrete network although our theoretical
considerations assume the real line as domain.} This rate-neuron
network constitutes an intermediate step towards a network of spiking
neurons. Each neuron has a fixed in-degree $K_{X}$ (fixed number
of incoming connections) per source population $X\in\mathrm{\{E,I}\}$
with connections selected randomly within a distance $R_{X}$. A normalization
of weights with the in-degree, $w'_{X}=w_{X}/K_{X}$, allows us to
interpret $p$ as a connection probability. The time constant $\tau$
and the delay $d$ are the same as in the neural-field model. As nonlinear
gain function in \prettyref{eq:intdiff} we choose $\psi\left(u\right)=\tanh\left(u\right)$.

\begin{figure}[t]
\begin{centering}
\includegraphics[width=0.85\textwidth]{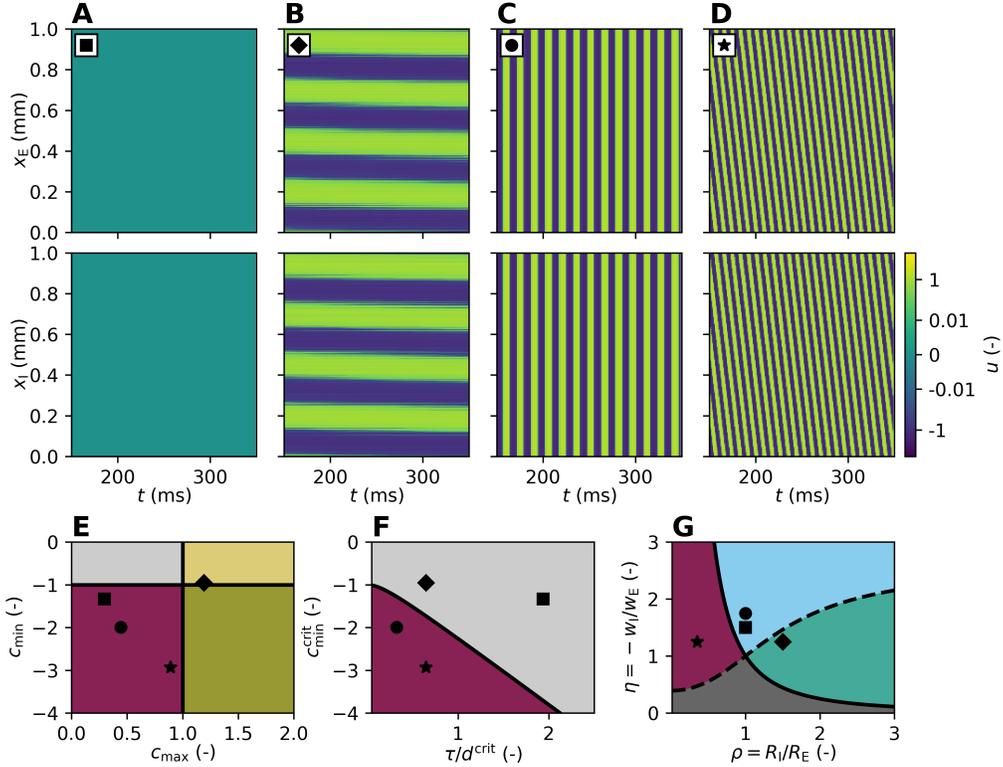}
\par\end{centering}
\centering{}\caption{\textbf{Predictions from linear stability analysis lead to spatiotemporal
patterns in simulated network of nonlinear rate neurons. }Different
parameter combinations, selected according to stability conditions
in \prettyref{tab:conditions}, cause pattern formation in rate-neuron
network with $\mathrm{tanh}$ gain function. \textbf{A–D}~Color-coded
activity per neuron over time. \textcolor{black}{Neurons are shown
at their position on the ring.}\textcolor{brown}{} \textbf{E–G}~Phase
diagrams showing conditions and parameter choices indicated by corresponding
markers. Purple regions indicate the possibility for wave trains.
\textbf{\textcolor{black}{E}}\textcolor{black}{~Color code indicates
stability based on minimum $c_{\mathrm{min}}$ and maximum $c_{\mathrm{max}}$.
Gray: Both $c_{\mathrm{min}}$ and $c_{\mathrm{max}}$ stable. Dirty
yellow: $c_{\mathrm{max}}$ unstable and $c_{\mathrm{min}}$ stable.
Dirty green: $c_{\mathrm{max}}$ unstable and $c_{\mathrm{min}}$
undetermined. Purple: $c_{\mathrm{max}}$ stable and $c_{\mathrm{min}}$
undetermined.}\textbf{ A}~Stable activity (square marker).\textbf{
B}~Spatial oscillations (diamond marker). \textbf{C}~Temporal oscillations
(circular marker). \textbf{D}~Wave trains (star marker). Parameters:
$d$, $R_{\mathrm{E}}$ and $R_{\mathrm{I}}$ as in \prettyref{fig:rasters_spatiotemporal_patterns}A–D,
$w_{\mathrm{E}}=2.73$ in all panels.\textbf{ A}~$w_{\mathrm{I}}=-4.10$.
\textbf{B}~$w_{\mathrm{I}}=-3.42$. \textbf{C}~$w_{\mathrm{I}}=-4.79$.
\textbf{D}~$w_{\mathrm{I}}=-3.42$. \label{fig:rate_simulation}
}
\end{figure}

The neuron activity of four rate-network simulations with different
parameter combinations are shown in \prettyref{fig:rate_simulation}A-D.
The location of the specific parameter combinations is illustrated
in \prettyref{fig:rate_simulation}E-G with corresponding markers
in the phase diagrams that visualize the stability conditions shown
in \prettyref{fig:conditions_pTW} derived with the neural-field model.
Wave trains are possible if parameters are in the purple regions of
the diagrams.

The system simulated in \prettyref{fig:rate_simulation}A is stable
according to the corresponding conditions. The square marker in the
lower panels shows that $c_{\mathrm{max}}<1$ (panel E), and although
$c_{\mathrm{min}}<-1$, the delay is small such that the system is
far away from the bifurcation (panel F). Indeed, the activity appears
to not exhibit any spatial or temporal structure.

\prettyref{fig:rate_simulation}B illustrates a case where $c_{\mathrm{max}}>1$
causes an instability (diamond marker in panel E). The Hopf bifurcation
is remote in the parameter space (panel F) and panel G ensures $k_{\mathrm{max}}>0$.
A simulation of the corresponding rate-model network again confirms
the predictions and exhibits stationary spatial oscillations (or periodic
bumps). \textcolor{black}{The predicted spatial frequency is $k_{\mathrm{max}}/\left(2\pi\right)\approx\unit[3.74]{mm^{-1}}$
and we expect $L\cdot k_{\mathrm{max}}/\left(2\pi\right)$ bumps to
emerge. In this finite-sized system with periodic boundary conditions,
the bumps are homogeneously distributed across the domain and the
wave numbers are integers; here we observe four stripes.}

\prettyref{fig:rate_simulation}C demonstrates temporal oscillations
at the parameter combination indicated by the circular marker. We
here choose $c_{\mathrm{max}}<1$ and $c_{\mathrm{min}}<-1$ (panel
E). The latter condition leads to an entire range of delays that are
beyond the bifurcation in panel F; we choose a delay slightly larger
than the critical delay, lying to the left of the bifurcation curve.
Inferred from panel G, $k_{\mathrm{min}}=0$ and, as expected from
the analytical prediction, the oscillations observed in simulations
of the rate-neuron network are purely temporal.\textcolor{black}{Based
on the temporal eigenvalue with the largest real part, we predict
a temporal frequency of $\operatorname{Im}\left[\lambda_{\mathrm{min}}\right]/\left(2\pi\right)\approx\unit[66.68]{Hz}$
which fits well to the simulated oscillation frequency.}

Finally, \prettyref{fig:rate_simulation}D depicts wave trains (denoted
by star marker), as predicted by the analytically tractable neural-field
model. The instability results from $c_{\mathrm{min}}<c_{\mathrm{min}}^{\mathrm{crit}}$
(panel F) and occurs at $k_{\mathrm{min}}>0$ (panel G) while $c_{\mathrm{max}}$
remains stable (panel E). \textcolor{black}{With a spatial frequency
of $k_{\mathrm{min}}/\left(2\pi\right)\approx\unit[3.02]{mm^{-1}}$
and a temporal frequency of $\operatorname{Im}\left[\lambda_{\mathrm{min}}\right]/\left(2\pi\right)\approx\unit[121.01]{Hz}$,
the predicted wave-propagation speed is $\operatorname{Im}\left[\lambda_{\mathrm{min}}\right]/\left(k_{\mathrm{min}}\right)\approx\unit[0.04]{mm/ms}$
which is in agreement with the simulated propagation speed of the
wave train.}

\subsection{Linearization of spiking network model\label{sub:Linearization-of-spiking}}

To assess the validity of the predictions obtained from the analytical
model for biologically more realistic spiking-neuron networks, we
next linearize the dynamics of spiking leaky integrate-and-fire (LIF)
neurons and derive a linear system similar to the neural-field model
above. The sub-threshold dynamics of a single LIF neuron $i$ with
exponentially decaying synaptic currents is described by a set of
differential equations for the time evolution of the membrane potential
$V_{i}$ and its synaptic current $I_{i}$ as

\begin{equation}
\begin{split}\tau_{\mathrm{m}}\frac{\partial V_{i}}{\mathrm{\partial}t} & =-V_{i}+I_{i}\left(t\right),\\
\tau_{\mathrm{s}}\frac{\partial I}{\mathrm{\partial}t} & =-I_{i}+\tau_{\mathrm{m}}\sum_{j}J_{ij}s_{j}\left(t-d\right),
\end{split}
\label{eq:subthreshold_LIF}
\end{equation}
where we follow the convention of \cite{FourcaudTrocme05_311} (see
\prettyref{eq:lif_dynamics_phy} in \nameref{sec:Methods} for the
relation to physical units). This definition, with both quantities
$V_{i}$ and $I_{i}$ having the same unit, conserves the total integrated
charge per impulse flowing into the membrane independent of the choice
of the synaptic time constant $\tau_{\mathrm{s}}$. The membrane time
constant, defined as $\tau_{\mathrm{m}}=R_{\mathrm{m}}C_{\mathrm{m}}$
with membrane resistance $R_{\mathrm{m}}$ and membrane capacitance
$C_{\mathrm{m}}$, couples current to the capacitance. We here assume
$\tau_{\mathrm{s}}$ to be much smaller than $\tau_{\mathrm{m}}$.
The term $s_{j}\left(t\right)=\sum_{k}\delta\left(t-t_{k}^{j}\right)$
denotes a spike train of neuron $j$ which is connected to neuron
$i$ with a constant connection strength $J_{ij}$ and transmission
delay $d$. Whenever $V_{i}$ reaches the threshold $V_{\theta}$,
a spike is emitted and the membrane potential is reset to the resting
potential $V_{\mathrm{r}}$ and voltage-clamped for the refractory
period $\tau_{\mathrm{ref}}$.

\textcolor{brown}{}\textcolor{black}{We now assume that, conditioned
on the time-dependent spike emission rate $\nu_{i}(t)$ of neuron
$i$, spikes are generated independently, thus with Poisson statistics
(see, e.g., \cite[Section 3.5]{Brunel99} for a discussion of this
approximation). A neuron then receives a superposition of many such
uncorrelated and Poisson-distributed input spikes, so that the probability
distribution $p(V,I,t)$ follows a Chapman-Kolmogorov equation. }\textcolor{black}{We
further assume the amplitudes of postsynaptic potentials to be small,
and perform a Kramers-Moyal expansion \cite{Ricciardi99,Risken96}
up to second order, which yields a Fokker-Planck equation for $p(V,I,t)$
in which the first and second infinitesimal moments appear as}

\textcolor{black}{
\begin{equation}
\begin{split}\mu_{i}\left(t\right) & =\tau_{\mathrm{m}}\sum_{j}J_{ij}\,\nu_{j}\left(t-d\right),\\
\sigma_{i}^{2}\left(t\right) & =\tau_{\mathrm{m}}\sum_{j}J_{ij}^{2}\,\nu_{j}\left(t-d\right).
\end{split}
{\color{red}}\label{eq:mean_var-3}
\end{equation}
}\textcolor{black}{Here $(\mu_{i},\sigma_{i})$ can be thought of as
the first two moments of a Gaussian white noise in the diffusion approximation
\cite{Tuckwell88b,Amit-1997_373,Ricciardi99}. This synaptic noise
in the input to neuron $i$ depends on the receiving neuron index
$i$ and hence on its position. Therefore $\mu_{i}$ and $\sigma_{i}$
also depend on the column $J_{ij}\forall j$ of the connectivity matrix.}\textcolor{brown}{}
Such a mean-field approach has been employed previously to study networks
of spiking neurons without spatial structure \cite{Amit-1997_373,Brunel99,Brunel00_183,Lindner05_061919}.,
where $\nu_{j}=\nu$ are identical for all $j$, given by the population-averaged
firing rate.

\textcolor{brown}{}\textcolor{black}{We extend on this approach by
assuming that the neurons are placed uniformly with density $\rho_{x}$
on a one-dimensional domain and apply the established procedure to
obtain a continuum limit \cite{Wilson1973}: A volume element $\mathrm{d}x$
of the one-dimensional domain contains the number $\rho_{x}\mathrm{d}x$
of neurons. We further assume that an incoming connection from a neuron
at position $y$ to a neuron at position $x$ is drawn independently
and identically distributed (i.i.d.) with probability proportional
to a spatial profile $\widetilde{p}(x-y)$. Hence the $J_{ij}$ in
\prettyref{eq:mean_var-3} are i.i.d. Bernoulli variables that take
the value $J$ with a probability $\propto\widetilde{p}(x-y)$ and
are zero otherwise. The expressions for the first and second infinitesimal
moment in \prettyref{eq:mean_var-3} of a neuron at position $x$
under expectation of the random connectivity are then}

\textcolor{black}{
\begin{equation}
\begin{split}\mu\left(x,t\right) & =\tau_{\mathrm{m}}J\int_{-\infty}^{\infty}\,\widetilde{p}\left(x-y\right)\,\nu\left(y,t-d\right)\,\rho_{x}\mathrm{d}y,\\
\sigma^{2}\left(x,t\right) & =\tau_{\mathrm{m}}J^{2}\int_{-\infty}^{\infty}\,\widetilde{p}\left(x-y\right)\,\nu\left(y,t-d\right)\,\rho_{x}\mathrm{d}y.
\end{split}
\label{eq:mean_var-1}
\end{equation}
We find it convenient to introduce a normalized profile $p\left(x-y\right)=\frac{\widetilde{p}\left(x-y\right)}{\int\widetilde{p}\left(x^{\prime}\right)\,\mathrm{d}x^{\prime}}$
and to define the number of incoming connections per neuron as $K:=\int\widetilde{p}\left(x^{\prime}\right)\,\rho_{x}\,\mathrm{d}x^{\prime}$.}\textcolor{brown}{}

\textcolor{black}{In the following we formally write down an evolution
equation for the rate $\nu\left(x,t\right)$. We denote as $\nu\left(x,t\right)=F\left[\mu\left(x,\circ\right),\,\sigma\left(x,\circ\right)\right]\left(t\right)$
the firing rate of a LIF neuron at position $x$ at time $t$ described
by \prettyref{eq:subthreshold_LIF} that is driven by a white noise
with mean $\mu(x,t)$ and variance $\sigma^{2}\left(x,t\right)$.
Clearly, $F\left[\mu,\sigma\right]\left(t\right)$ is a causal functional
of its two arguments, which are functions of time (temporal argument
denoted by $\circ$). The firing rate of the neuron at time point
$t$ only depends on the statistics of its input up to this point,
hence on $\mu(x,s)\,\forall s\le t$ and $\sigma(x,s)\,\forall s\le t$.
Thus we can define this functional as $F\left[\mu\left(\circ\right),\sigma\left(\circ\right))\right]\left(t\right):=\langle\delta\left(t-t_{k}\right)\rangle_{\xi}$,
where $t_{k}$ are the time-points of the threshold crossings of \prettyref{eq:subthreshold_LIF}
under expectation $\left\langle \right\rangle _{\xi}$ over the realization
of the white noise $\xi$ with moments \prettyref{eq:mean_var-1}
and $\delta$ is the Dirac distribution. Since the statistics of the
input, the functions $\mu\left(t\right)$ and $\sigma\left(t\right)$,
are direct functions of the firing rate $\nu\left(y,t-d\right)$ by
\prettyref{eq:mean_var-1}, the evolution equation takes the form
\begin{align}
\begin{split} & \nu\left(x,t\right)\\
 & =F\left[\tau_{\mathrm{m}}JK\int_{-\infty}^{\infty}\,p\left(x-y\right)\,D_{d}\,\nu\left(y\right)\,\mathrm{d}y,\,\tau_{\mathrm{m}}J^{2}K\int_{-\infty}^{\infty}\,p\left(x-y\right)\,D_{d}\,\nu\left(y\right)\,\mathrm{d}y\right]\left(t\right),
\end{split}
\label{eq:evolution_equation}
\end{align}
where the delay operator $D_{d}$ is defined to act on the second
(temporal) argument of the function as $[D_{d}\nu(x)](t)=\nu(x,t-d)$.
In principle, the functional $F$ can be computed – for example, by
solving the mean first-passage time for the membrane potential $V$
to exceed the threshold. For that purpose, we would drive the neuron
with a Gaussian noise with a given, time-dependent statistics parameterized
by $\mu$ and $\sigma^{2}$. Powerful numerical methods are available
for this purpose \cite{Richardson07_021919}. For the purpose of the
present work, however, we do not need to determine $F$ in complete
generality, since we are only interested in a linear stability analysis
of a spatially and temporally homogeneous state $\nu\left(x,t\right)=\nu_{0}$.
Hence it is sufficient to study the stability of \prettyref{eq:evolution_equation}
with respect to spatio-temporal deviations of the form}

\textcolor{black}{
\begin{equation}
\nu\left(x,t\right)=\nu_{0}+\delta\nu\left(x,t\right),\qquad\delta\nu\ll\nu_{0}.
\end{equation}
Linearizing \prettyref{eq:evolution_equation} we obtain (by a functional
Taylor expansion or Volterra expansion to first order)}

\textcolor{black}{
\begin{align}
\begin{split} & \nu_{0}+\delta\nu\left(x,t\right)\\
 & =F\left[\mu_{0},\sigma_{0}^{2}\right]+\int_{-\infty}^{\infty}\,p\left(x-y\right)\,\int_{-\infty}^{t}\,h_{\nu}(\mu_{0},\sigma_{0},t-s)\,\delta\nu\left(y,s-d\right)\,\mathrm{d}s\,\mathrm{d}y+\mathcal{O}\left(\delta\nu^{2}\right),\\
 & \text{with }h_{\nu}(\mu_{0},\sigma_{0},t-s)=\tau_{\mathrm{m}}JK\,\frac{\delta F\left[\mu_{0},\sigma_{0}^{2}\right]\left(t\right)}{\delta\mu\left(s\right)}+\tau_{\mathrm{m}}J^{2}K\,\frac{\delta F\left[\mu_{0},\sigma_{0}^{2}\right]\left(t\right)}{\delta\sigma^{2}\left(s\right)},
\end{split}
\label{eq:linearization}
\end{align}
where we introduce the short hand $\mu_{0}=\tau_{\mathrm{m}}JK\nu_{0}$
and $\sigma_{0}^{2}=\tau_{\mathrm{m}}J^{2}K\nu_{0}$. With $\nu_{0}=F\left[\mu_{0},\sigma_{0}^{2}\right]$
the first line of \prettyref{eq:linearization} cancels the corresponding
term on the left hand side and we obtain a linear convolution equation
for the rate deflection $\delta\nu$, whose spectral properties we
need to analyze. The stationary firing rate $\nu_{0}$ can be determined
self-consistently from this condition (see \prettyref{eq:siegert}
in \nameref{sec:Methods}). The functional derivatives 
\begin{align*}
\frac{\delta F\left[\mu_{0},\sigma_{0}^{2}\right]\left(t\right)}{\delta\mu\left(s\right)} & \equiv\lim_{\epsilon\to0}\,\frac{1}{\epsilon}\big(F\left[\mu_{0}+\epsilon\,\delta\left(\circ-s\right),\,\sigma_{0}^{2}\right]\left(t\right)-F\left[\mu_{0},\sigma_{0}^{2}\right]\left(t\right)\big),
\end{align*}
(and analogous for $\delta F/\delta\sigma^{2}$) are, by the right
hand side of this definition, the responses of the system with respect
to an impulse-like perturbation of $\mu$ and $\sigma^{2}$, respectively.}\textcolor{brown}{{}
}\textcolor{black}{We denote these as
\begin{equation}
\begin{split}h_{\mu}\left(t-s\right) & \equiv\frac{\delta F\left[\mu_{0},\sigma_{0}^{2}\right]\left(t\right)}{\delta\mu\left(s\right)},\\
h_{\sigma^{2}}\left(t-s\right) & \equiv\frac{\delta F\left[\mu_{0},\sigma_{0}^{2}\right]\left(t\right)}{\delta\sigma^{2}\left(s\right)},
\end{split}
\label{eq:response_kernels}
\end{equation}
which are causal functions of $t-s$ only, since we linearize around
a time-translation invariant state and causality clearly requires
both kernels to vanish for $s>t$. These functions can analytically
be computed in the Fourier domain for LIF models with instantaneous
synapses \cite{Brunel99,Lindner01_2934}, for fast colored noise \cite{Schuecker15_transferfunction},
and in the adiabatic limit for slow synapses \cite{Morenobote06_028101,MorenoBote10_1528};
see \prettyref{subsec:Linearization-of-the} for details. The form
of the response kernels \prettyref{eq:response_kernels} is given
in Eqs \ref{eq:Jannis}–\ref{eq:Fourier_Laplace} in \nameref{sec:Methods}.
These expressions are obtained by a perturbative calculation on the
level of the Fokker-Planck equation that is correct to leading order
in $\mathcal{O}(\sqrt{\tau_{s}/\tau_{m}})$ \cite{Schuecker15_transferfunction},
thus constitude good approximations for sufficiently short synaptic
time constants}.\textcolor{brown}{{} }\textcolor{black}{With this notation,
the linearized dynamics \prettyref{eq:linearization} obeys a convolution
equation in space and time}

\textcolor{black}{
\begin{equation}
\delta\nu\left(x,t\right)=\int_{-\infty}^{\infty}\,p\left(x-y\right)\,\int_{-\infty}^{t}\,h_{\nu}(t-s)\,\delta\nu(y,s-d)\,\mathrm{d}y\,\mathrm{d}s,\label{eq:delta_nu}
\end{equation}
whose stability properties can be analyzed in Fourier domain by standard
methods.}\textcolor{brown}{}

In the following section we will ignore the kernel $h_{\sigma^{2}}$,
because its contribution is usually small \cite{Schuecker15_transferfunction}.
\prettyref{eq:delta_nu} provides a linearized system for the spiking
model that is continuous in space and time and enables a direct comparison
with the neural-field model in the following section.

\subsection{Comparison of neural-field and spiking models\label{sub:Comparison-of-models}}

The linearization of the LIF model presented in the preceding section
is the analogue to taking the derivative $\psi^{\prime}$ of the gain
function in the linear stability analysis of the neural-field model
in \prettyref{sub:Linear-stability-analysis}. \textcolor{black}{By
the assumption of conditional independence of spike trains given their
firings rate, we achieve that the state of the spiking network is
described by the time-dependent firing rate profile $\nu(x,t)$. Its
temporal evolution follows \prettyref{eq:evolution_equation}. This
function therefore conceptually plays the same role as $u(x,t)$ in
the neural-field model}\textcolor{brown}{.} Therefore the results
for the neural field model carry over to the spiking case. To expose
the similarities between the linearized systems of the spiking model
and the neural-field model, we may bring the equations for the deviation
from baseline activity 
\begin{align}
\delta o(x,t) & =\begin{cases}
\delta u(x,t) & \text{neural field}\\
\delta\nu(x,t) & \text{spiking}
\end{cases}
\end{align}
to the form of the convolution equation

\begin{equation}
\begin{split}\delta o(x,t) & =\left[h\ast\delta i\right](x,t)\\
\delta i\left(x,t\right) & =\int_{-\infty}^{\infty}p\left(x-y\right)\,\delta o\left(y,t-d\right)\,\mathrm{d}y,
\end{split}
\label{eq:lin_eq_s_r}
\end{equation}
where the only difference is the convolution kernel relating the deviation
from the input $\delta i$ to those of the output $\delta o$ defined
as
\begin{align}
h(t) & :=\begin{cases}
h^{\mathrm{nf}}(t):=\Theta(t)\,\frac{w}{\tau}\,\mathrm{e}^{-\frac{t}{\tau}} & \text{neural field}\\
h^{\mathrm{s}}(t):=\tau_{\mathrm{m}}JK\,h_{\mu}(t) & \text{spiking.}
\end{cases}\label{eq:response_functions_rate_spiking}
\end{align}

The kernel on the first line is the fundamental solution (Green's
function) of the linear differential operator appearing on the left
hand side of \prettyref{eq:intdiff}, including the coupling weight
$w$. As a consequence, the characteristic equations for both models
result from the Fourier-Laplace ansatz $\delta o\left(x,t\right)=\mathrm{e}^{\mathrm{i}kx}\mathrm{e}^{\lambda t}$
which relates the eigenvalues $\lambda$ to the wave number $k$ as

\begin{equation}
H\left(\lambda\right)\cdot\mathrm{e}^{-\lambda d}\cdot\widehat{p}\left(k\right)=1.\label{eq:char_eq_s_r}
\end{equation}
\textcolor{black}{The effective transfer function $H$ describes the
linear input-output relationship \prettyref{eq:lin_eq_s_r} in the
Laplace domain. It is obtained as the Laplace transform of \prettyref{eq:response_functions_rate_spiking}
of the respective functions for the spiking model $h^{\mathrm{s}}\left(t\right)$
and for the neural-field model $h^{\mathrm{nf}}$$\left(t\right)$.}
As a result we obtain the transfer function for the neural-field model
\begin{align}
H^{\mathrm{nf}}\left(\lambda\right) & =\frac{1}{1+\lambda\tau}w.\label{eq:transfer_rate}
\end{align}
The corresponding expression for the effective spiking transfer function
$H^{\mathrm{s}}\left(\lambda\right)$ results from Eqs \ref{eq:Jannis}-\ref{eq:Fourier_Laplace}
in \nameref{sec:Methods}.

\subsubsection{Parameter mapping\label{sub:Parameter-mapping}}

\begin{figure}[t]
\begin{centering}
\includegraphics[width=0.85\textwidth]{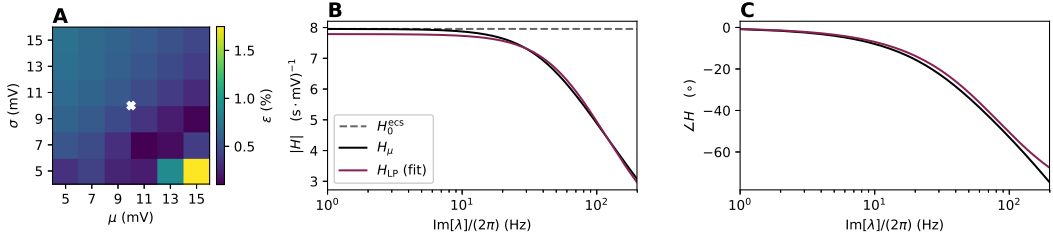}
\par\end{centering}
\centering{}\caption{\textbf{Transfer function of spiking neuron model and its approximation.
A~}Fitting error of the low-pass filter approximation of the transfer
function for LIF neurons derived in \cite{Schuecker15_transferfunction}
over $\mu$ and $\sigma$ (given relative to the reset potential).
The fitting error $\epsilon=\sqrt{\epsilon_{\tau}^{2}+\epsilon_{H_{0}}^{2}}$
is color-coded.\textbf{ B}~Amplitude of the transfer function and
approximation (legend). Dashed line illustrates $H_{0}$ following
from the analytically-determined effective coupling strength (see
\prettyref{eq:effective_coupling_strength} in \nameref{sec:Methods}).
\textbf{C}~Phase. The white cross in panel A indicates the working
point $\left(\mu,\sigma\right)$ selected for the transfer function
shown in panels B and C and used in the simulations throughout the
study. \label{fig:transfer_function_fit}}
\end{figure}

\textcolor{black}{So far the stability analysis shows that the characteristic
equations for both the neural-field and the spiking model have the
same form \prettyref{eq:char_eq_s_r} given a proper definition of
the respective transfer functions. The transfer function characterizes
the transmission of a small fluctuation in the input to the output
of the neuron model. Because the transfer functions differ between
the two models, it is a priori unclear whether their characteristic
equations have qualitatively similar solutions. To provide evidence
that this is indeed the case, in the following we devise a procedure
that identifies solutions of the characteristic equations in \prettyref{eq:char_eq_s_r}
for the rate model and the spiking model, and develop a practical
method to obtain one solution from the other.}

\textcolor{black}{To this end we use that the transfer function of the
LIF model in the fluctuation-driven regime investigated here can be
approximated by a first order low-pass (LP) filter \cite{Lindner01_2934,Brunel01_2186,Helias13_023002}
(see in particular} \textcolor{black}{\cite[Fig 1]{Brunel01_2186})
with effective parameters $H_{0}$ and $\tau$}

\textcolor{black}{
\begin{equation}
H_{\mu}\left(\lambda\right)\approx H_{\mathrm{LP}}\left(\lambda\right)=\frac{H_{0}}{1+\lambda\tau},
\end{equation}
where $H_{\mu}$ is the Fourier transform of $h_{\mu}$, defined in
\prettyref{eq:response_kernels}.}\textcolor{brown}{{} }This simplified
transfer function is similar to the the transfer function \prettyref{eq:transfer_rate}
of the neural-field model, and thereby relates the phenomenological
parameters $w$ and $\tau$ of the neural-field model to the biophysically
motivated parameters of the spiking model.

\textcolor{brown}{}\textcolor{black}{We perform a least squares fit
between $H_{\mathrm{LP}}\left(\lambda\right)$ and $H_{\mu}\left(\lambda\right)$
to obtain the values for the parameters $\tau$ and $H_{0}$. According
to \prettyref{eq:response_functions_rate_spiking}, $H_{0}$ directly
relates to $w$ as}

\textcolor{black}{
\begin{equation}
w=H_{0}\tau_{\mathrm{m}}JK,\label{eq:w_H0}
\end{equation}
which follows by noting that $\int h_{\mu}(t)=H_{\mu}(0)\approx H_{0}$.}
The goodness of the fit of this transfer function to the first-order
low-pass filter depends on the mean $\mu$ and variance $\sigma$
of the synaptic input, as shown in \prettyref{fig:transfer_function_fit}A.
The color-coded error of the fit combines the relative errors from
both fitting parameters: $\epsilon=\sqrt{\epsilon_{\tau}^{2}+\epsilon_{H_{0}}^{2}}$.
For the majority of working points $\left(\mu,\sigma\right)$ the
error is $<1\%$ but the relative errors increase abruptly towards
the mean-driven regime. In this regime input fluctuations are small
and the mean input predominantly drives the membrane potential towards
threshold, so that the model fires regularly and the transfer function
exhibits a peak close to the firing frequency \cite{Lindner01_2934,Brunel01_2186}.
We here fix the working point to the parameters indicated by the white
cross (see \prettyref{eq:external_rates} in \nameref{sec:Methods})
for all populations, resulting in a common effective time constant
$\tau$. Here, we obtain a time constant $\tau=\unit[1.94]{ms}$ which
thus lies in between the synaptic time constant, $\tau_{\mathrm{s}}=\unit[0.5]{ms}$,
and the membrane time constant, $\tau_{\mathrm{m}}=\unit[5]{ms}$,
of the LIF neuron model. For these parameters, \prettyref{fig:transfer_function_fit}B
shows the amplitude and \prettyref{fig:transfer_function_fit}C the
phase of the original transfer function $H_{\mu}\left(\lambda\right)$
in black and the fitted transfer function $H_{\mathrm{LP}}\left(\lambda\right)$
in purple. The dashed gray line denotes $H_{0}$ obtained by computing
the effective coupling strength from linear response theory, $H_{0}^{\mathrm{ecs}}$,
as a reference (see \prettyref{eq:effective_coupling_strength} in
\nameref{sec:Methods}).

\subsubsection{Linear interpolation between the transfer functions }

\begin{figure}[t]
\begin{centering}
\includegraphics[width=0.85\textwidth]{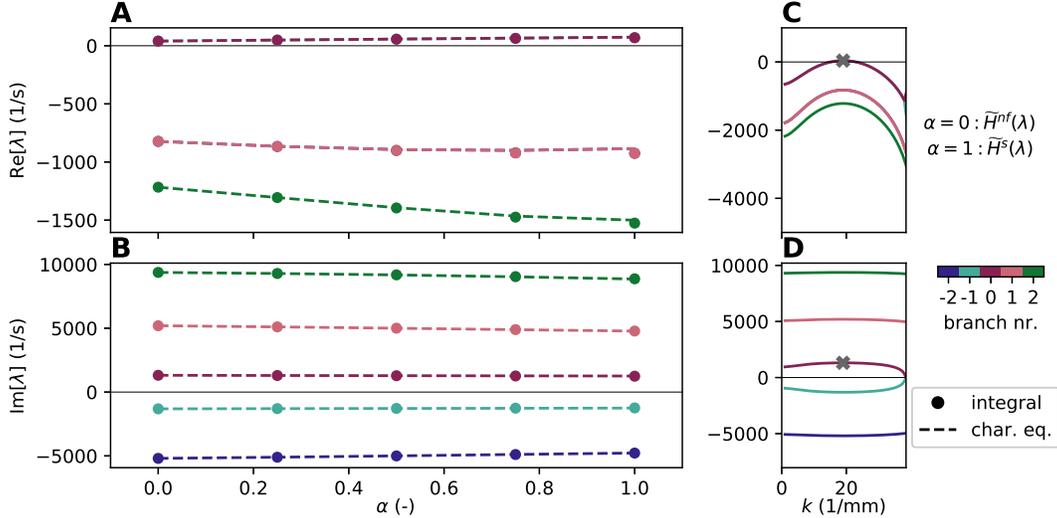}
\par\end{centering}
\centering{}\caption{\textbf{Linear interpolation between neural-field $\left(\alpha=0\right)$
and spiking $\left(\alpha=1\right)$ model for eigenvalue close to
bifurcation.} \textbf{A}~Real and \textbf{B~}imaginary part of the
eigenvalue $\lambda$ as a function of the linear interpolation parameter
$\alpha$ for the characteristic equation in \prettyref{eq:comb_char_eq}.
The solution at $\alpha=0$ for the neural-field model is exact.\textbf{
C}~Real and \textbf{D}~imaginary part of the eigenvalues (same units
but different scaling as in A and B) with analytically exact solution
(by Lambert W function, $\alpha=0$) as functions of the wave number
$k$. Different branches $b$ are color-coded (legend); \textbf{$b=0$
}corresponds to the principal branch with the maximum real eigenvalue
(gray cross). Circular markers denote the linear interpolation according
to the numerical integration of \prettyref{eq:lin_interpol_integral}.
Dashed line segments for the linear interpolation are obtained by
solving the characteristic equation \prettyref{eq:comb_char_eq} numerically.
Both are evaluated at the same values for $\alpha$. Parameters: $d=\unit[1.5]{ms}$,
$R_{\mathrm{E}}=\unit[0.2]{mm}$, $R_{\mathrm{I}}=\unit[0.07]{mm}$,
$g=5.$ \label{fig:linear_interpolation} }
\end{figure}

Evaluating the characteristic equation for the neural-field model
yields an exact solution for each branch of the Lambert W function,
given by \prettyref{eq:eigenvalues_rate_LW}. For this model we already
established that the principle branch is the most unstable one. An
equivalent condition is not known for the general response kernel
of the LIF neuron. To asses whether we may transfer the result for
the neural-field model to the spiking case, we investigate the correspondence
between the two characteristic equations that are both of the form
\prettyref{eq:char_eq_s_r} but with different transfer functions.
For this purpose, we define an effective transfer function

\begin{equation}
H_{\alpha}\left(\lambda\right)=\alpha H^{\mathrm{s}}\left(\lambda\right)+\left(1-\alpha\right)H^{\mathrm{nf}}\left(\lambda\right),\label{eq:eff_tf_lambda_of_alpha}
\end{equation}
with the parameter $\alpha$ that linearly interpolates between the
effective transfer functions of the spiking and the neural-field model:
$H_{\alpha=0}\left(\lambda\right)=H^{\mathrm{nf}}$$\left(\lambda\right)$
and $H_{\alpha=1}\left(\lambda\right)=H^{\mathrm{s}}$$\left(\lambda\right)$.
\prettyref{fig:linear_interpolation} illustrates two different ways
for solving the combined characteristic equation

\begin{equation}
H_{\alpha}\left(\lambda\right)\cdot\mathrm{e}^{-\lambda d}\cdot\widehat{p}\left(k\right)=1.\label{eq:comb_char_eq}
\end{equation}
The first results from computing the derivative $\partial\lambda/\partial\alpha$
(see Eqs \ref{eq:derivative_implicit_fct}-\ref{eq:R_lambda} in \nameref{sec:Methods})
from the combined characteristic equation and integrating numerically
with the exact solution of the neural-field model at $\alpha=0$ for
each branch $b$ as initial condition:

\begin{equation}
\lambda\left(\alpha\right)=\int_{0}^{\alpha}\frac{\mathrm{\partial}\lambda}{\partial\alpha^{'}}\,\mathrm{d}\alpha^{'},\qquad\lambda\left(0\right)=\lambda_{b}\label{eq:lin_interpol_integral}
\end{equation}
with

\begin{equation}
\frac{\mathrm{\partial}\lambda}{\mathrm{\partial}\alpha}=-\frac{H^{\mathrm{s}}\left(\lambda\right)-H^{\mathrm{nf}}\left(\lambda\right)}{\alpha\frac{\partial H^{\mathrm{s}}\left(\lambda\right)}{\partial\lambda}+\left(1-\alpha\right)\frac{\partial H^{\mathrm{nf}}\left(\lambda\right)}{\partial\lambda}-d\cdot H_{\alpha}\left(\lambda\right)}.\label{eq:d_lambda_d_alpha}
\end{equation}

The spatial profile only enters the initial condition, and the derivative
\prettyref{eq:d_lambda_d_alpha} is independent of the wave number
$k$.

As an alternative approach, we directly solve the combined characteristic
equation \prettyref{eq:comb_char_eq} numerically with the known initial
condition. \prettyref{fig:linear_interpolation}A and B indicate that
only the principle branch ($b=0$) becomes positive while the other
branches remain stable. The branches come in complex conjugate pairs.
For the numerical solution of the characteristic equation, we fix
the wave number to the value of $k$ that corresponds to the maximum
real eigenvalue. 

The analysis shows that we may ignore the danger of branch crossing
since different branches remain clearly separated in \prettyref{fig:linear_interpolation}A
and B. In addition, the eigenvalue on the principle branch is mostly
independent of $\alpha$, even if the system is close to the bifurcation
(when the real part of $\lambda_{0}$ is close to zero). Thus for
all values of $\alpha$ we expect qualitatively similar bifurcations,
including $\alpha=1$. This justification transfers the rigorous results
from the bifurcation analysis of the neural-field model in \prettyref{sub:Conditions-for-traveling}
and \prettyref{sub:Application-to-a-concrete-network}, and corresponding
effective parameters, to the spiking model.

\subsection{Validation by simulation of spiking neural network\label{sub:Validation-in-simulation}}

The \nameref{sec:Introduction} illustrates spatiotemporal patterns
emerging in a spiking network simulation in \prettyref{fig:rasters_spatiotemporal_patterns}
and the subsequent sections derive a theory describing the mechanisms
underlying such patterns. Finally, the parameter mapping between the
spiking and the neural-field model explains the origin of the spike
patterns by transferring the conditions found for the abstract neural-field
model in \prettyref{sub:Conditions-for-traveling} and \prettyref{sub:Application-to-a-concrete-network}
to the spiking case. This section validates that the correspondence
between network parameters in the two models is not incidental but
covers the full phase diagram.

\begin{figure}[H]
\begin{centering}
\includegraphics[width=0.85\textwidth]{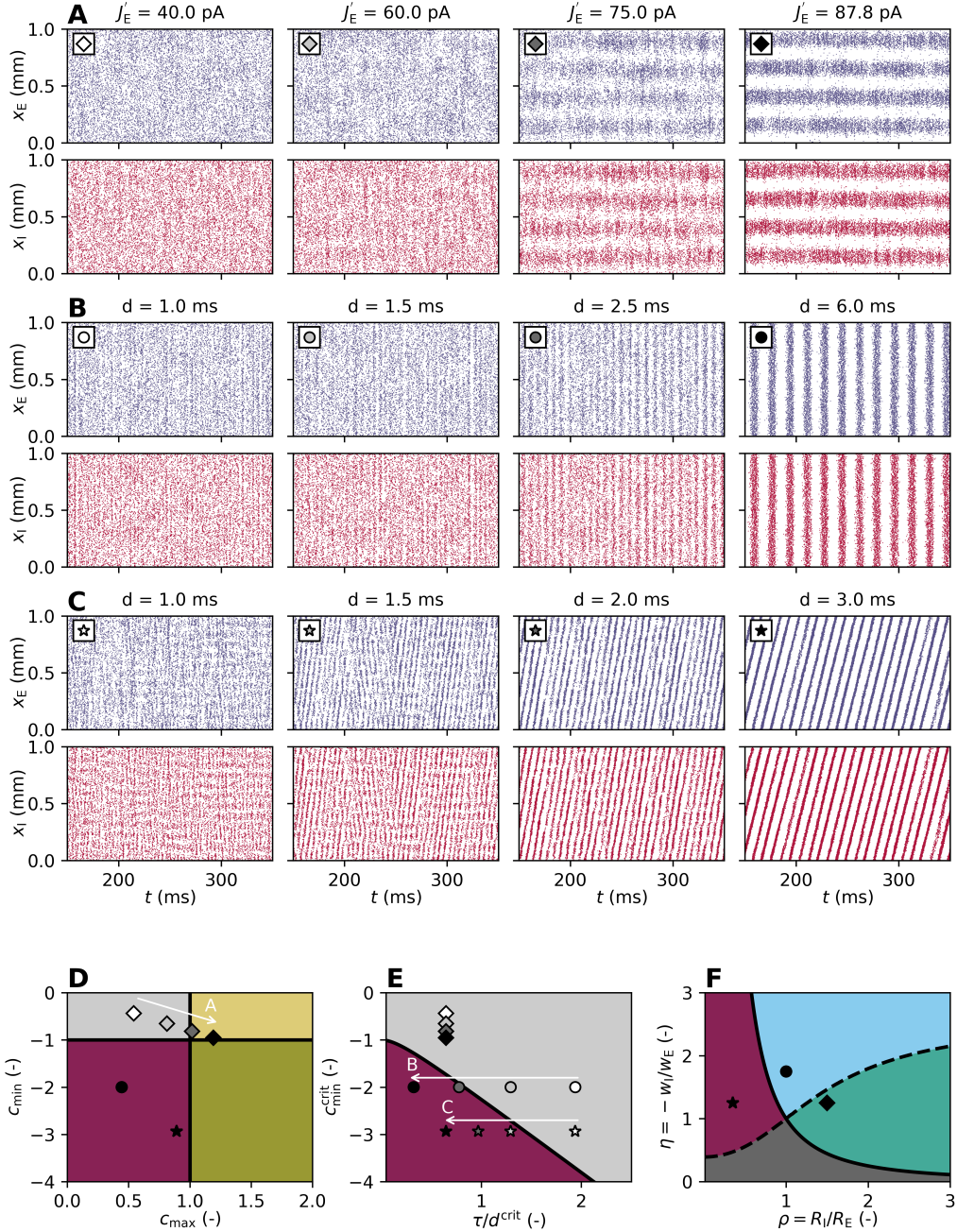}
\par\end{centering}

\caption{\textbf{Transitions from theoretically stable states to spatiotemporal
patterns in spiking network simulation. A–C}~Spike rasters showing
transition to network states in \prettyref{fig:rasters_spatiotemporal_patterns}B–D
(same markers, same parameter combinations). The changed parameter
value is given on top of each raster plot. \textbf{A~}Increasing
recurrent weight $J_{\mathrm{E}}^{'}$ leads to onset of spatial oscillations.
\textbf{B}~Increasing synaptic delay $d$ leads to onset of temporal
oscillations at $k=0$. \textbf{C}~Increasing delay $d$ leads to
onset of temporal oscillations at $k>0$, i.e., wave trains. \textbf{D–E}~Gray
shaded markers and white arrows labeled according to respective panel
A-C in phase diagrams indicate sequences of parameter combinations
and breakdown of stability at $c_{\mathrm{max}}=1$ or at $c_{\mathrm{min}}=c_{\mathrm{min}}^{\mathrm{crit}}$.
For each sequence in panels A–C, delay $d$, excitatory profile width
$R_{\mathrm{E}}$, inhibitory profile width $R_{\mathrm{I}}$, and
the relative synaptic strength $g$ correspond to the values given
in \prettyref{fig:rasters_spatiotemporal_patterns}B–D with corresponding
markers. \label{fig:transitions}}
\end{figure}

In the following, we simulate a network with the same neural populations
and spatial connectivity used in the nonlinear rate-network in \prettyref{fig:rate_simulation},
but replace the rate-model neurons by spiking neurons, and map the
parameters as described in \prettyref{sub:Parameter-mapping}. The
network model characterizes all neurons by the same working point
(see \prettyref{eq:external_rates} in \nameref{sec:Methods}), which
means that the connectivity matrix for the excitatory-inhibitory network
has equal rows; entries in \prettyref{eq:connectivity_rate} depend
on the presynaptic population alone. Therefore the relative in-degree
$\gamma=K_{\mathrm{I}}/K_{\mathrm{E}}$ and the relative synaptic
strength $g=-J_{\mathrm{I}}/J_{\mathrm{E}}$ parametrize the spiking-network
connectivity matrix as

\begin{equation}
M\left(r\right)=\tau_{m}J_{\mathrm{E}}K_{\mathrm{E}}\,\left(\begin{array}{cc}
p_{\mathrm{E}}\left(r\right) & -\gamma g\,p_{\mathrm{I}}\left(r\right)\\
p_{\mathrm{E}}\left(r\right) & -\gamma g\,p_{\mathrm{I}}\left(r\right)
\end{array}\right).\label{eq:connectivity_spiking_EI}
\end{equation}
The rightmost panels of \prettyref{fig:transitions}A–C show the same
simulation results as \prettyref{fig:rasters_spatiotemporal_patterns}B–D;
likewise the panels of \prettyref{fig:rasters_spatiotemporal_patterns}
have parameters that correspond to those of the rate-neuron network
in \prettyref{fig:rate_simulation}. The different patterns in \prettyref{fig:rasters_spatiotemporal_patterns}B–D
emerge by gradually shifting a single network parameter that switches
the system from a stable state (white filled markers in \prettyref{fig:transitions}D
and E), across intermediate states (gray-scale filled markers) to
the final states where stability is lost and the patterns have formed
(black filled markers). Arrows visualize the sequences in the phase
diagrams \prettyref{fig:transitions}D and E and the markers reappear
in the upper left corners of the corresponding raster plots in \prettyref{fig:transitions}A–C.

The sequence of panels in \prettyref{fig:transitions}A illustrates
a gradual transition from a stable (AI) state to spatial oscillations
attained by increasing the amplitudes of excitatory postsynaptic current
(PSC) amplitudes $J_{\mathrm{E}}^{'}$ in the network. With $J^{'}$
we denote the weight as a jump in current while $J$ denotes a jump
in voltage in the physical sense, and the relationship is: $J^{'}=C_{\mathrm{m}}J/\tau_{s}$
(see \prettyref{eq:lif_dynamics_phy} in \nameref{sec:Methods}).
The parameter variation thus homogeneously scales the effective profile
$c$ but preserves the shape of the reduced profile $\widetilde{c}$
(fixed position of diamond marker in panel F). Simultaneously an increasing
rate of the external Poisson input compensates for the reduced PSC
amplitudes to maintain the fixed working point $\left(\mu,\sigma\right)$
of the neurons (see \prettyref{eq:external_rates} in \nameref{sec:Methods}).
Diamond markers in \prettyref{fig:transitions}D show that along its
path the system crosses the critical value $c_{\mathrm{max}}=1$,
while $c_{\mathrm{min}}>c_{\mathrm{min}}^{\mathrm{crit}}(\tau/d^{\mathrm{crit}})$
stays in the stable regime, as shown in panel E. However, even for
$c_{\mathrm{max}}\lesssim1$ (for $J_{E}^{'}=\unit[60]{pA}$) the
network activity already exhibits weak spatial oscillations. 

Choosing the synaptic delay $d$ as a bifurcation parameter highlights
the onset of temporal oscillations for the case $k=0$ (panel B sequence,
circular markers) and spatiotemporal oscillations for the case $k>0$
(sequence in \prettyref{fig:transitions}C, star markers). In contrast
to the case of purely spatial waves in panel A, the procedure preserves
the effective spatial profile (fixed positions in panels D and F)
and the system crosses the transition curve in panel E due to increasing
delay alone, thus decreasing the ratio $\tau/d$.

\prettyref{fig:transitions}C illustrates the gradual transition to
wave trains, where $c_{\mathrm{max}}$ remains in the theoretically
stable regime at all times, but is close to the critical value of
$1$ (see the star marker in panel D). As a result, we observe spatial
oscillations with a spatial frequency given by $k_{\mathrm{max}}$
before and even after the Hopf bifurcation. For delays longer than
the critical delay, mixed states occur in which different instabilities
due to $c_{\mathrm{max}}$ and $c_{\mathrm{min}}$ compete. \textcolor{black}{The
different spatial frequencies $k_{\mathrm{max}}/\left(2\pi\right)$
and $k_{\mathrm{min}}/\left(2\pi\right)$ become visible.} For delay
values well past the bifurcation, this mixed state is lost resulting
in a dependency only on $c_{\mathrm{min}}$ and wave trains with a
spatial frequency that depends on $k_{\mathrm{min}}$.

\section{Discussion\label{sec:Discussion}}

The present study employs mean-field theory \cite{Brunel99} to rigorously
map a spiking network model of leaky integrate-and-fire (LIF) neurons
with constant transmission delay to a neural-field model. We use a
conceptually similar linearization as Kriener et al. \cite{Kriener14}
combined with an analytical expressions for the transfer function
in the presence of colored synaptic noise \cite{Schuecker15_transferfunction}.
The insight that this transfer function in the fluctuation-driven
regime resembles the one of a simple first-order low-pass filter facilitates
the parameter mapping between the two models. The resulting analytically
tractable effective rate model depends on the dynamical working point
of the spiking network that is characterized by both the mean and
the variance of the synaptic input. By means of bifurcation theory,
in particular linear Turing instability analysis \cite{Coombes05,Coombes07_51901,Venkov07_1},
we investigate the origin of spatiotemporal patterns such as temporal
and spatial oscillations and in particular wave trains emerging in
spiking activity. The mechanism underlying these waves encompasses
delay-induced fast global oscillations, as described by Brunel and
Hakim \cite{Brunel99}, with spatial oscillations due to a distance-dependent
effective connectivity profile. We derive analytical conditions for
pattern formation that are exclusively based on general characteristics
of the effective connectivity profile and the delay. The profile is
split into a static weight that is either excitatory or inhibitory
for a given neural population, and a spatial modulation that can be
interpreted as a distance-dependent connection probability. Given
the biological constraint that connection probabilities depend on
distance but weights do not, wave trains cannot occur in a single
homogeneous population irrespective of the shape of distance-dependent
connection probability. Only the effective connectivity profile of
two populations (excitatory and inhibitory), permits solutions where
a mode with finite non-zero wave number is the most unstable one,
a prerequisite for the emergence of nontrivial spatial patterns such
as wave trains. We therefore establish a relation between the anatomically
measurable connectivity structure and observable patterns in spiking
activity. The predictions of the analytically tractable neural-field
model are validated by means of simulations of nonlinear rate-unit
networks \cite{Hahne17_34} and of networks composed of LIF-model
neurons, both using the same simulation framework \cite{Nest2140}.
In our experience, the ability to switch from a model class with continuous
real-valued interaction to a model class with pulse-coupling by changing
a few lines in the formal high-level model description increases the
efficiency and reliability of the research.

The presented mathematical correspondence between these a priori
distinct classes of models for neural activity has several implications.
First, as demonstrated by the application in the current work, it
facilitates the transfer of results from the well-studied domain of
neural-field models to spiking models. The insight thus allows the
community to arrive at a coherent view of network phenomena that appear
robustly and independently of the chosen model. Second, the quantitative
mapping of the spiking model to an effective rate model in particular
reduces the parameters of the former to the set of fewer parameters
of the latter; single-neuron and network parameters are reduced to
just a weight and a time constant. This dimensionality reduction of
the parameter space conversely implies that entire manifolds of spiking
models are equivalent with respect to their bifurcations. Such a reduction
supports systematic data integration: Assume a researcher wants to
construct a spiking model that reproduces a certain spatiotemporal
pattern. The presented expressions permit the scientist to restrict
further investigations to the manifold in parameter space in line
with these observations. Variations of parameters within this manifold
may lead to phenomena beyond the predictions of the initial bifurcation
analysis. Additional constraints, such as firing rates, degree of
irregularity, or correlations, can then further reduce the set of
admissible parameters.

To keep the focus on the transferability of results from a neural-field
to a spiking model, the present study restricts the analysis to a
rather simple network model. In many cases, extensions to more realistic
settings are straight forward. As an example, we perform our analysis
in one-dimensional space. In two dimensions, the wave number becomes
a vector and bifurcations to periodic patterns in time and space can
be constructed (see \cite[Section 8.4]{Ermentrout98b} and \cite{Coombes05}).
Likewise, we restricted ourselves to a constant synaptic delay like
Roxin et al. \cite{Roxin05,Roxin06} because it enables a separation
of a spatial component, the shape of the spatial profile, and a temporal
component, the delay. A natural next step is the inclusion of an axonal
distance-dependent delay term as for instance in \cite{Hutt03_351}
to study the interplay of both delay contributions \cite{Veltz13_1566}.
For simplification, we use here a boxcar-shaped spatial connectivity
profile in the demonstrated application of our approach. For the emergence
of spatiotemporal patterns, however, the same conditions on the connectivity
structure and the delays hold for more realistic exponentially decaying
or Gaussian-shaped profiles \cite{Hellwig00_111,Perin11,Schnepel15_3818}.
If the spatial connectivity profiles are monotonically decaying in
the Fourier domain (as it is the case for exponential or Gaussian
shapes), the Fourier transform of the effective profile of a network
composed of an excitatory and an inhibitory population exhibits
at most one zero-crossing. Either the minimum or the maximum are attained
at a non-zero and finite wave number $k$, but not both. With a cosine-shaped
effective profile, only a single wave number dominates by construction
\cite{Roxin05,Roxin06}. Here, we decided for the boxcar shape because
of its oscillating Fourier transform that allows us to study competition
between two spatial frequencies corresponding to the two extrema.

Similar to our approach, previous neural-field studies describe the
spatial connectivity profile as a symmetric probability density function
(see, for example, \cite{Wyller07_75}). For our aim, to establish
a link to networks of discrete neurons, the interpretation as a connection
probability and the separation from a weight are a crucial addition.
This assumption enables us to distinguish between different neural
populations, to analyze the shape of the profile based on parameters
for the excitatory and the inhibitory contribution, and to introduce
biophysically motivated parameters for the synaptic strength. Starting
directly with an effective profile that includes both, excitation
and inhibition, such as (inverse) Mexican hat connectivity, is mathematically
equivalent and a common approach in the neural-fields literature \cite{Hutt03_351,Atay05_644,Coombes05,Roxin05}.
But it neglects the biological separation of neurons into excitatory
and inhibitory populations according to their effect on postsynaptic
targets (Dale's law \cite{Eccles54_524}) and their different spatial
reach of connectivity \cite{Stepanyants09_3555}. A result of this
simplification, these models can produce waves even with a single
homogeneous population \cite{Roxin05,Atay06_670,Venkov07_1}, while
with homogeneous stationary external drive we show that at least
two populations are required.

Local excitation and distant inhibition are often used to support
stationary patterns such as bumps, while local inhibition and distant
excitation are associated with non-stationary patterns such as traveling
waves \cite{Cross93_851,Ermentrout98b,Hutt03_351}. For sufficiently
long synaptic delays, we also observe wave trains with local inhibition
and distant excitation, as often observed in cortex \cite{Stepanyants09_3555}.
However, we show that the reason for this is the specific shape of
the effective spatial profile, and not only the spatial reach itself.
Our argumentation is therefore in line with Hutt et al. \cite{Hutt05_30,Hutt08_541}
who demonstrate that wave instabilities can even occur with local
excitation and distant inhibition for specific spatial interactions.
The spatial connectivity structure and related possible activity states
are in addition important factors for computational performance or
function of model networks \cite{Legenstein07_323,Pyle17_18103}.

\textcolor{black}{In \prettyref{sub:Network-simulation-with-nonlinear-rate},
we compute the spatial and temporal oscillation frequencies as well
as the wave-propagation speed. Such quantities are directly comparable
with experimental measurements. The conduction speed in unmyelinated
fibers is in the $\unit[0.1]{mm/ms}$ range and the propagation speed
of mesoscopic waves is of similar order of magnitude \cite{Girard01,Muller18_255}.
Our prediction based on the current choice of model parameters is
with $\unit[0.04]{mm/ms}$ in the same order of magnitude.}

The parameter mapping between a neural-field and a spiking model in
this study relies on the insight that the transfer function of the
LIF neuron in the fluctuation-driven regime resembles the one of a
simple first-order low-pass filter. Since this approximation not only
holds for LIF neurons, but also for other spiking neuron models, our
results are transferable. A further candidate model with this property
is the exponential integrate-and-fire model \cite{Fermani15_cns_P206}.
Other examples include Nordlie et al. \cite{Nordlie10} who characterize
the firing-rate responses of LIF neurons with strong alpha-shaped
synaptic currents and similarly Heiberg et al. \cite{Heiberg13_359}
for a LIF neuron model with conductance based synapses and potassium-mediated
afterhyperpolarization currents proposed previously \cite{Casti08_235}.

In the literature, the time constant of neural-field models is often
associated with the membrane or the synaptic time constant \cite{Bressloff00_169,Bressloff14_8,Pyle17_18103}.
Here, we observe that the time constant of the neural-field model
derived from the network of spiking neurons falls in between the two.
In line with \cite{Gerstner00,Nordlie10}, we suggest to reconsider
the meaning of the time constant in neural-field models.

A limitation of the approach employed here is that the linear theory
is only exact at the onset of waves. Beyond the bifurcation, it is
possible that nonlinearities in the spiking model govern the dynamics
and lead to different prevailing wave numbers or wave frequencies
than predicted. Roxin et al. \cite{Roxin06} report that the stability
of traveling waves depends crucially on the nonlinearity. Nevertheless
they do not observe traveling waves in their spiking-network simulations.
In the present work, however, we identify biophysically motivated
neuron and network parameters that allow wave trains to establish
in a spiking network. Still, we had to increase the delay beyond the
predicted bifurcation point to obtain a stable wave pattern.

Furthermore, the theory underlying the mapping of the spiking network
to the neural-field model is based on the diffusion approximation
and therefore only applicable for sufficiently small synaptic weights.
Widely distributed synaptic weights, for example, may lead to larger
deviations. We here primarily target a wave-generating mechanism for
cortical networks. Since in other brain regions involved neuron types,
connectivity structures and input characteristics are different, other
mechanisms for pattern formation not covered in this work need to
be taken into account \cite{Muller12_222}.

The working-point dependence of the neural-field models derived here
offers a new interpretation of propagating activity measured in vivo
\cite{Takahashi15,Denker18_1}. Even if the anatomical connectivity
remains unchanged during a period of observation, the stability of
the neural system can be temporarily altered due to changes in activity.
The transfer function of a LIF neuron depends on the mean and the
variance of its input, and we have shown that stability is related
to its parametrization. In particular, local changes of activity,
for example due to a spatially confined external input, can affect
stability and hence influence whether a signal remains rather local
or travels across the cortical surface. That means, we would relate
the tendency of a neural network to exhibit spatiotemporal patterns
not only to its connectivity, but also to its activity state that
can change over time.

\section{Appendix\label{sec:Methods}}

\subsection{Linear stability analysis}

\subsubsection{Derivation of the characteristic equation}

With the Fourier-Laplace ansatz $u\left(x,t\right)=e^{\mathrm{i}kx}\mathrm{e}^{\lambda t}$
for the integro-differential equation in \prettyref{eq:intdiff} linearized
around $u_{0}$ and the choice to set the slope of the gain function
to unity, the characteristic equation in \prettyref{eq:char_eq} results
from

\begin{equation}
\begin{split}\tau\lambda\,\mathrm{e}^{ikx}\mathrm{e}^{\lambda t} & =-\mathrm{e}^{ikx}\mathrm{e}^{\lambda t}+\int_{-\infty}^{\infty}wp\left(x-y\right)\mathrm{e}^{iky}\mathrm{e}^{\lambda\left(t-d\right)}\,\mathrm{d}y\\
\tau\lambda & =-1+w\mathrm{e}^{-\lambda d}\int_{-\infty}^{\infty}p\left(x-y\right)\mathrm{e}^{-ik\left(x-y\right)}\,\mathrm{d}y\\
 & =-1-w\mathrm{e}^{-\lambda d}\int_{\infty}^{-\infty}p\left(r\right)\mathrm{e}^{-ikr}\,\mathrm{d}r,\qquad r=x-y\\
 & =-1+w\mathrm{e}^{-\lambda d}\underbrace{\int_{-\infty}^{\infty}p\left(r\right)\mathrm{e}^{-ikr}\,\mathrm{d}r}_{\equiv\widehat{p}\left(k\right)}.
\end{split}
\label{eq:derivation_char_eq}
\end{equation}
In the last row, we recognize the Fourier transform $\widehat{p}$
of the spatial profile $p$.\\

\subsubsection{Effective connectivity profile for two populations}

While the connectivity $m$ is a scalar in the one-population model,
it is a matrix $M$ in the case of two populations (given in \prettyref{eq:connectivity_rate}).
The ansatz for deriving the characteristic equation in the latter
case reads $\delta u\left(x,t\right)=v\mathrm{e}^{\mathrm{i}kx}\mathrm{e}^{\lambda t}$,
with $v$ denoting a vector of constants. This leads to the auxiliary
eigenvalue problem

\begin{equation}
c\left(k\right)\,v=\widehat{M}\left(k\right)\,v,\label{eq:aux_eig_prob}
\end{equation}
where $c$ denotes an eigenvalue and $\widehat{M}$ is an auxiliary
matrix containing the Fourier transforms of the entries of $M$:

\begin{equation}
\widehat{M}\left(k\right)=\left(\begin{array}{cc}
w_{\mathrm{EE}}\,\widehat{p}_{\mathrm{EE}}\left(k\right) & w_{\mathrm{EI}}\,\widehat{p}_{\mathrm{EI}}\left(k\right)\\
w_{\mathrm{IE}}\,\widehat{p}_{\mathrm{IE}}\left(k\right) & w_{\mathrm{II}}\,\widehat{p}_{\mathrm{II}}\left(k\right)
\end{array}\right).
\end{equation}
\prettyref{eq:aux_eig_prob} possesses a nontrivial solution $v$
if and only if $\det\left(\widehat{M}\left(k\right)-c\left(k\right)\mathbbm{1}\right)=0$.
\prettyref{eq:eigenvalues_P_hat} explicitly states the two eigenvalues
$c_{1,2}$ solving this equation. These eigenvalues constitute the
effective profile in the characteristic equation in \prettyref{eq:char_eq}
that hence holds also for the two-population case.

\subsubsection{Largest real part on principle branch of Lambert W function}

The function $x\left(W\right)=W\,\mathrm{e}^{W}$ has a minimum at
$W=-1$, no real solution for $x<-\mathrm{e}^{-1}$, a single solution
for $x>0$, and two solutions for $x\in[-e^{-1},0)$. Typically, the
term ‘principal branch’ of the Lambert W function with branch number
$b=0$ refers to the real branch defined on the interval $[-e^{-1},\infty)$,
where for negative arguments the larger solution is considered. Here
we extend the definition to the whole real line by the complex branch
with maximal real part and positive imaginary part on $(-\infty,-e^{-1})$.

We demonstrate that the branch of the Lambert W function with the
largest real part is the principal branch. Considering only real-valued
arguments $x\in\mathbb{R}$, we write $W\left(x\right)=\left|W\left(x\right)\right|\mathrm{e}^{i\varphi}=\alpha+i\beta$
and

\begin{align}
W\left(x\right)\mathrm{e}^{W\left(x\right)} & =\left|W\left(x\right)\right|\,\mathrm{e}^{\alpha}\,\mathrm{e}^{i\left(\varphi+\beta\right)}=x\in\mathbb{R}\label{eq:largest_real_part_0}\\
\rightarrow\mathrm{e}^{i(\varphi+\beta)} & =\pm1,\label{eq:largest_real_part_1}
\end{align}
where $\varphi\in\left[-\pi,\pi\right]$ is the principal value. We
index the branches by $q\in\mathbb{Z}$ according to the number of
half-cycles of the exponential in \prettyref{eq:largest_real_part_1}:
$\varphi+\beta=q\cdot\pi$. The branch number is equal to $b=\left\lfloor \frac{q}{2}\right\rfloor $
with $\left\lfloor \cdot\right\rfloor $ denoting the floor function.
The principle branch is therefore given by the index $q=0$ for $x\ge0$
and by $q=1$ for $x<0.$

Taking the absolute square of \prettyref{eq:largest_real_part_0}
yields the real equation

\begin{equation}
x^{2}\,\mathrm{e}^{-2\alpha}=\alpha^{2}+\beta^{2}.\label{eq:largest_real_part_2}
\end{equation}
Without loss of generality we may assume $\beta\geq0$; this is certainly
true for the real solutions with $\beta=0$ and it also holds for
one of the complex solutions for any complex pair. Complex solutions
come in conjugate pairs due to the symmetry $(\varphi,\beta)\to(-\varphi,-\beta)$
exhibited by \prettyref{eq:largest_real_part_1} and \prettyref{eq:largest_real_part_2}.
Since each member of a pair has by definition the same real part,
it is sufficient to consider only the member with positive imaginary
part $\beta>0$.

To prove that the real part $\alpha$ of $W$ is maximal for $b=0$,
we show that $\alpha$ is a decreasing function of $\beta$ along
the solutions of \prettyref{eq:largest_real_part_0}. Investigating
the intersections of the left-hand side and the right-hand side of
\prettyref{eq:largest_real_part_2} as a function of $\alpha$ illustrates
how increasing the imaginary part $\beta$ affects the real part $\alpha$.
The left-hand side is a decaying function of $\alpha$ with an intercept
of $x^{2}$. The right-hand-side is a parabola with an offset of $\beta^{2}$.

For $x\in(-\infty,-e^{-1})\cup\left[0,\infty\right)$, an intersection
occurs either at a positive real part $\alpha\geq0$ if $x^{2}\geq\beta^{2}$,
or at a negative real part $\alpha<0$ if $x^{2}<\beta^{2}$. Increasing
$\beta$ moves the parabola upwards and therefore the intersection
to the left, meaning that $\alpha$ decreases with increasing $\beta$.

For $x\in[-e^{-1},0)$, we distinguish the cases $\beta=0$ and $\beta>0$
which both have only solutions with $\alpha<0$. First, the two real
solutions ($q=\pm1$) existing in this interval correspond to two
simultaneously occurring intersections; in addition a third intersection
is created by the squaring \prettyref{eq:largest_real_part_2} but
it is not an actual solution of \prettyref{eq:largest_real_part_0}.
The intersection at the larger real part per definition corresponds
to the principal branch with index $q=1$. Second, the complex solutions
are indexed by odd numbers $q$ with $\left|q\right|>1$. Taking into
account the interval where $\varphi$ is defined, the imaginary part
is bounded from below such that $\beta\geq2\pi$ for non-principal
branches. Analogous to the previously discussed interval of $x$,
there exists only one intersection between the exponential function
and the parabola for large values of $\beta$ (in particular: $x^{2}<\beta^{2}$)
that moves towards smaller values of $\alpha$ with increasing $\beta$.

So in summary we have shown that for real $x$, the principal branch
harbors the solutions with maximal real part $\alpha$.

\subsubsection{Characteristic equation with Lambert W function}

The characteristic equation in \prettyref{eq:char_eq} can be rewritten
in terms of the Lambert W function to \prettyref{eq:eigenvalues_rate_LW}
using the transformation:

\begin{equation}
\begin{split}\left(1+\tau\lambda\right)\mathrm{e}^{\lambda d} & =c\left(k\right)\,|\cdot\frac{d}{\tau}\mathrm{e}^{\frac{d}{\tau}}\\
\left(d\lambda+\frac{d}{\tau}\right)\mathrm{e}^{d\lambda+\frac{d}{\tau}} & =c\left(k\right)\frac{d}{\tau}\mathrm{e}^{\frac{d}{\tau}}\\
d\lambda+\frac{d}{\tau} & =W\left(c\left(k\right)\frac{d}{\tau}\mathrm{e}^{\frac{d}{\tau}}\right).
\end{split}
\label{eq:char_eq_to_LW}
\end{equation}
The last step collects terms using the definition of the Lambert
W function, $z=W\left(z\right)\mathrm{e}^{W\left(z\right)}$with $z\in\mathbb{C}$.

\subsection{Properties of the spatial profile}

We assume that the spatial profile $p$ is a symmetric probability
density function, which implies that its Fourier transform $\widehat{p}$,
also called the characteristic function, is real valued and even.
Further, we can prove that $\widehat{p}\in(-1,1]$ and that $\widehat{p}$
attains $1$ only at the origin in two steps:
\begin{itemize}
\item $|\widehat{p}(k)|\leq1\;\text{ for all }k\in\mathbb{R}$:
\end{itemize}
\begin{equation}
\begin{split}|\widehat{p}(k)| & =\left|\int_{-\infty}^{\infty}p(r)\mathrm{e}^{-ikr}\,\mathrm{d}r\right|\leq\int_{-\infty}^{\infty}\left|p(r)\mathrm{e}^{-ikr}\right|\,\mathrm{d}r\\
 & =\int_{-\infty}^{\infty}p(r)\,\mathrm{d}r=1\quad\text{ for all }k\in\mathbb{R},
\end{split}
\label{eq:abs_p_hat1}
\end{equation}

\begin{itemize}
\item $|\widehat{p}(k)|<1\;\text{ for all }k\neq0$:
\end{itemize}
\begin{equation}
\begin{split}\left|\int_{-\infty}^{\infty}p(r)\mathrm{e}^{-ikr}\mathrm{d}r\right| & \leq\int_{-\infty}^{\infty}p(r)\left|\cos\left(kr\right)\right|\mathrm{\,d}r\\
 & <\int_{-\infty}^{\infty}p(r)\mathrm{\,d}r=1\quad\text{ for all }k\neq0,
\end{split}
\label{eq:abs_p_hat2}
\end{equation}
\textcolor{black}{because $\left|\cos\left(kr\right)\right|<1$ except
for a set of measure zero}\textcolor{black}{{} in $r$ if $k\neq0$,
that does not influence the value of the integral.}

\subsection{Transition curves for reduced profile}

\begin{figure}[t]
\begin{centering}
\includegraphics[width=0.85\textwidth]{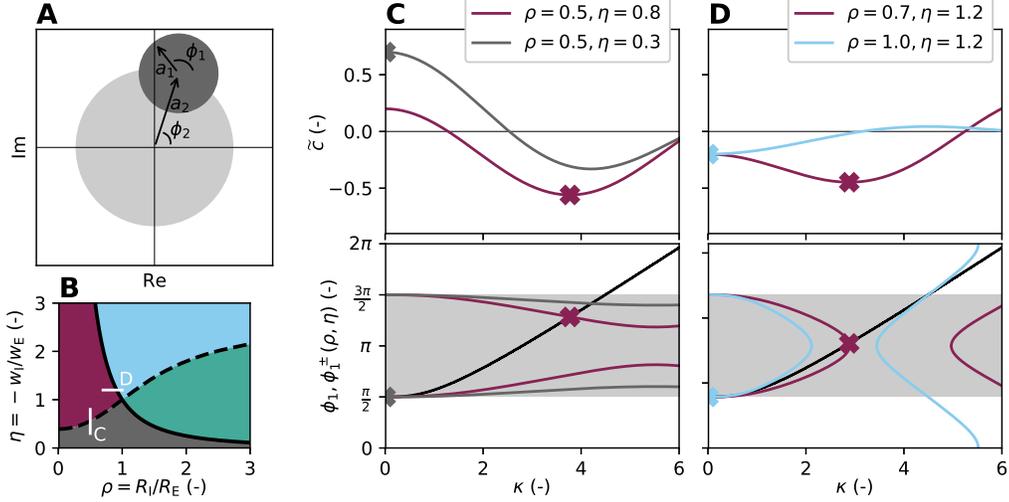}
\par\end{centering}

\caption{\textbf{Graphical analysis for extrema of reduced profile for derivation
of transition curves.} \textbf{A}~The condition for the extremum
\prettyref{eq:deriv_B} amounts to the addition of two vectors in
the complex plane whose sum is purely imaginary. The vectors have
lengths $a_{1}$ and $a_{2}$ and angles $\phi_{1}$ and $\phi_{2}$,
defined in \prettyref{eq:abs_phase}. \textbf{B~}Diagram of \prettyref{fig:conditions_pTW}F
with indicated parameter combinations $\left(\rho,\eta\right)$ as
used in panels C and D. \textbf{C–D}~Reduced profile $\widetilde{c}$
(top) and $\phi_{1}$ and $\phi_{1}^{\pm}$ from \prettyref{eq:phi1_pm}
vs. $\kappa$ (bottom) for two different combinations of $\left(\rho,\eta\right)$
with line colors corresponding to regions in panel B. \textbf{C}~$\left|\widetilde{c}_{\mathrm{min}}\right|>\widetilde{c}_{\mathrm{max}}$
in purple and vice versa in dark gray. \textbf{D}~$\widetilde{c}_{\mathrm{min}}$
at $\kappa=0$ in light blue and at $\kappa>0$ in purple.\label{fig:graphical_approach_transition_lines}}
\end{figure}

We here use a graphical approach to derive the transition curves shown
first in \prettyref{fig:conditions_pTW}F. A necessary condition for
an extreme value of the reduced profile $\widetilde{c}\left(\kappa\right)$
from \prettyref{eq:reduced_effective_profile} located at $\kappa^{\ast}$
is: $\begin{array}{rcl}
\frac{\partial}{\partial\kappa}\widetilde{c}\left(\kappa\right)|_{\kappa^{\ast}} & = & 0.\end{array}$ With the derivative 

\begin{equation}
\frac{\partial}{\partial\kappa}\widetilde{c}\left(\kappa\right)=\frac{\cos\left(\kappa\right)}{\kappa}-\frac{\sin\left(\kappa\right)}{\kappa^{2}}-\eta\frac{\cos\left(\rho\kappa\right)}{\kappa}+\eta\frac{\sin\left(\rho\kappa\right)}{\rho\kappa^{2}},\label{eq:deriv_B_full}
\end{equation}
this condition can be rewritten as

\begin{equation}
\begin{split}0 & =\operatorname{Re}\left[\left(\kappa+i\right)\mathrm{e}^{i\kappa}-\frac{\eta}{r}\left(\rho\kappa+i\right)\mathrm{e}^{i\rho\kappa}\right]\\
 & =\operatorname{Re}\left[a_{1}\mathrm{e}^{i\phi_{1}}+a_{2}\mathrm{\mathrm{e}}^{i\phi_{2}}\right]\\
 & =a_{1}\cos\left(\phi_{1}\right)+a_{2}\cos\left(\phi_{2}\right),
\end{split}
\label{eq:deriv_B}
\end{equation}
where $a_{1}$ and $a_{2}$ are the absolute values of the complex
numbers and $\phi_{1}$ and $\phi_{2}$ their phases, given by

\begin{equation}
\begin{split}a_{1}\left(\kappa\right) & =\sqrt{1+\kappa^{2}}\\
\phi_{1}\left(\kappa\right) & =\kappa+\frac{\pi}{2}-\arctan\left(\kappa\right)\\
a_{2}\left(\kappa;\rho,\gamma\right) & =\frac{\eta}{\rho}\sqrt{1+\rho^{2}\kappa^{2}}\\
\phi_{2}\left(\kappa;\rho\right) & =\rho\kappa+\frac{3\pi}{2}-\arctan\left(\rho\kappa\right).
\end{split}
\label{eq:abs_phase}
\end{equation}
The vanishing right-hand-side of \prettyref{eq:deriv_B} implies that
the term in the square brackets is purely imaginary. An example solution
for the case $a_{1}<a_{2}$ is illustrated in \prettyref{fig:graphical_approach_transition_lines}A
in the complex plane. Note that $a_{1}$ and $\phi_{1}$ are independent
of the parameters $\rho$ and $\eta$ in this representation. In our
graphical analysis, \prettyref{eq:deriv_B} is interpreted as the
sum of two vectors in the complex plane. As shown in \prettyref{fig:graphical_approach_transition_lines}A,
we determine $\phi_{1}$ as the angle at which the tip of the second
vector ends on the imaginary axis, which follows from elementary trigonometry
as 
\begin{equation}
\phi_{1}^{\pm}=\pi\pm\arccos\left(\frac{a_{2}}{a_{1}}\cos\left(\phi_{2}\right)\right).\label{eq:phi1_pm}
\end{equation}
The locations of extrema are then given by the intersections of $\phi_{1}^{\pm}$
with the second row of \prettyref{eq:abs_phase}. Here $\phi_{2}$
is determined from the last equation in \prettyref{eq:deriv_B}.

\prettyref{fig:graphical_approach_transition_lines}B reproduces \prettyref{fig:conditions_pTW}F.
The white bars connect points given by parameter combinations $\left(\rho,\eta\right)$
on both sides of the transition curves, and the parameters are specified
in panels C and D. The first transition curve $\eta_{t1}\left(\rho\right)$
(dashed line in \prettyref{fig:graphical_approach_transition_lines}B)
is determined by $\widetilde{c}_{\mathrm{max}}\left(\kappa_{\mathrm{max}}\right)=\left|\widetilde{c}_{\mathrm{min}}\left(\kappa_{\mathrm{min}}\right)\right|$,
that means it is determined by parameters $\left(\rho,\eta\right)$
for which the absolute values of the positive and negative extremum
of the profile are equal. The top panel of \prettyref{fig:graphical_approach_transition_lines}C
compares two reduced profiles obtained for a fixed value for $\rho$
and two values for $\eta$. The line colors correspond to the colored
regions in the diagram in \prettyref{fig:graphical_approach_transition_lines}B
for the respective parameter combination $\left|\widetilde{c}_{\mathrm{min}}\right|>\widetilde{c}_{\mathrm{max}}$
for the purple profile and vice versa for the dark gray profile. The
point with the maximum absolute value of each profile is indicated
with a cross. Exactly at the transition either $\kappa_{\mathrm{max}}$
or $\kappa_{\mathrm{min}}$ is zero (for example $\kappa_{0}=0$)
and the other one is non-zero (for example $\kappa_{1}>0$). This
condition, with \prettyref{eq:reduced_effective_profile}, yields
the absolute value for both extrema at the transition, where they
must be equal, thus $\left|\widetilde{c}\left(\kappa_{0}\right)\right|=\left|\widetilde{c}\left(\kappa_{1}\right)\right|=\left|1-\eta\right|$.
Any point on the transition curve is a unique triplet of parameters$\left(\rho,\eta,\kappa_{1}\right)$,
and with the condition $\frac{\partial}{\partial\kappa}\widetilde{c}\left(\kappa\right)|_{\kappa_{1}}=0$
we obtain two equations that need to be fulfilled at each point for
$\kappa=\kappa_{1}$:

\begin{equation}
\begin{split}1-\eta & =\frac{\sin\left(\kappa\right)}{\kappa}-\eta\frac{\sin\left(\rho\kappa\right)}{\rho\kappa}\\
1-\eta & =\cos\left(\kappa\right)-\eta\cos\left(\rho\kappa\right).
\end{split}
\label{eq:cond_eq_trans1}
\end{equation}
The lower equation is obtained by identifying $\widetilde{c}\left(\kappa\right)$
in its derivative in \prettyref{eq:deriv_B_full}. We solve both
equations with respect to $\eta$ and equate them to get

\begin{equation}
\frac{1}{\kappa}\sin\left(\kappa\right)\left[1+\cos\left(\rho\kappa\right)\right]-\frac{1}{\rho\kappa}\sin\left(\rho\kappa\right)\left[1+\cos\left(\kappa\right)\right]+\cos\left(\rho\kappa\right)-\cos\left(\kappa\right)=0.\label{eq:roots_kappa_rho}
\end{equation}
For a given value of $\rho$, we compute the roots of the left-hand-side
expression, which defines $\kappa(\rho)$. The bottom panel of \prettyref{fig:graphical_approach_transition_lines}C
shows $\phi_{1}$ from \prettyref{eq:abs_phase} as a black line and
$\phi_{1}^{\pm}$ from \prettyref{eq:phi1_pm} for the parameters
of the two effective profiles (same color coding as in the top panel).
The intersections corresponding to the relevant extrema are highlighted
by crosses. This visual analysis allows us to identify the interval
for $\kappa$ in which zero-crossings of the left-hand side of \prettyref{eq:roots_kappa_rho}
as a function of $\kappa$ can correspond to the extrema, that is
$\kappa\in\left(0,\,4.49341\right)$ where the lower limit corresponds
to $\phi_{1}=\frac{\pi}{2}$ and the upper limit to $\phi_{1}=\frac{3\pi}{2}$.
The zero-crossing at the smallest non-zero $\kappa$ indicates the
extremum at $\kappa_{1}$. Finally, the transition curve is given
by

\begin{equation}
\eta_{t1}\left(\rho\right)=\frac{1+\cos\left(\kappa\left(\rho\right)\right)}{1+\cos\left(\rho\kappa\left(\rho\right)\right)},\label{eq:transition_line1}
\end{equation}
where $\kappa(\rho)$ is given by the roots of \eqref{eq:roots_kappa_rho}.

The second transition curve $\eta_{t2}\left(\rho\right)$ (solid line
in \prettyref{fig:graphical_approach_transition_lines}B) indicates
whether the extremum with the largest absolute value occurs at $\kappa=0$
or at $\kappa>0$. \prettyref{fig:graphical_approach_transition_lines}D
shows in the top panel two reduced profiles for a fixed value of $\eta$,
but two values for $\rho$ such that the $\widetilde{c}_{\mathrm{min}}$
occurs once at $\kappa_{\mathrm{min}}=0$ (light blue as in \prettyref{fig:graphical_approach_transition_lines}B)
and once at $\kappa_{\mathrm{min}}>0$ (purple as in \prettyref{fig:graphical_approach_transition_lines}B),
indicated by cross markers.

Graphical analysis using the bottom panel of \prettyref{fig:graphical_approach_transition_lines}D
indicates that this transition happens when $\phi_{1}^{-}$ at $\kappa\gtrsim0$
switches from lying slightly above (light blue line) to below (purple
line) the parameter-independent function $\phi_{1}$ (black line).
We observe that decreasing $\rho$ moves the intersection point and
with it the location of the extremum up the black line, starting from
$\kappa=0$ to larger values for $\kappa$. 

Close to the transition, the intersection point comes arbitrarily
close to $\kappa=0$, which permits local analysis by a Taylor expansion
of $\phi_{1}$ for small $\kappa$:

\begin{align}
\phi_{1}\left(\kappa\right) & \approx\frac{\pi}{2}+\frac{\kappa^{3}}{3}+\mathcal{O}\left(\kappa^{5}\right)\\
\phi_{1}^{-}\left(\kappa;\rho,\eta\right) & \approx\frac{\pi}{2}+\frac{\eta\rho\kappa^{3}}{3}+\mathcal{O}\left(\left(\rho\kappa\right)^{5}\right).
\end{align}
A comparison of the coefficients of the third-order polynomials then
gives the transition curve

\begin{equation}
\eta_{t2}\left(\rho\right)=\frac{1}{\rho^{2}},\label{eq:transition_line2}
\end{equation}
because this coefficient decides for small $\kappa$ whether $\phi_{1}$
(black line) or $\phi_{1}^{-}$ as a function of the parameters $\left(\rho,\eta\right)$
has a larger slope and lies on top.

\subsection{Linearization of the spiking model\label{subsec:Linearization-of-the}}

\subsubsection{Fast synaptic noise}

The stationary firing rate of a LIF neuron model subject to fast synaptic
noise has been derived in \cite{Brunel01_2186,Fourcaud02}. The linear
response of the model to time-dependent stimuli has been derived in
\cite[Eq 29]{Schuecker15_transferfunction}, by application of a general
reduction technique to a white noise system with displaced boundary
conditions.

\subsubsection*{Stationary firing rate}

The stationary firing rate $\nu_{0}$ in the limit of short synaptic
time constants ($\tau_{\mathrm{s}}\ll\tau_{\mathrm{m}}$) is given
by \cite[Eq A.1]{Fourcaud02,Helias13_023002}:

\begin{equation}
\begin{split}\nu_{0}^{-1} & =\tau_{\mathrm{r}}+\tau_{\mathrm{m}}\sqrt{\pi}\left(F\left(y_{\theta}\right)-F\left(y_{\mathrm{r}}\right)\right)\\
f\left(y\right) & =\mathrm{e}^{y^{2}}\left(1+\mathrm{erf}\left(y\right)\right),\qquad F\left(y\right)=\int^{y}f\left(y\right)dy\\
\mathrm{\mbox{with}\quad}y_{\left\{ \theta,r\right\} } & =\frac{V_{\left\{ \theta,r\right\} }-\mu}{\sigma}+\frac{\beta}{2}\sqrt{\frac{\tau_{\mathrm{s}}}{\tau_{\mathrm{m}}}},\qquad\beta=\sqrt{2}\left|\zeta\left(\frac{1}{2}\right)\right|,
\end{split}
\label{eq:siegert}
\end{equation}
where $\zeta$ denotes the Riemann's zeta function \cite{Abramowitz74}.

\subsubsection*{Transfer function\label{sub:Transfer-function}}

The linear response of the firing rate is described by the transfer
function, here denoted by $H_{\mu}$, that relates the modulation
of the firing rate $\delta\nu(\omega)$ to the modulation of the mean
$\delta\mu(\omega)$ as
\begin{align*}
\delta\nu(\omega) & =H_{G}(\omega)\,\delta\mu(\omega)+o(\delta\mu^{2}).
\end{align*}
It is computed based on the first term of \cite[Eq 29]{Schuecker15_transferfunction}

\begin{equation}
H_{G}\left(\omega\right)=\frac{\nu_{0}\frac{\sqrt{2}}{\sigma}}{1+i\omega\tau_{\mathrm{m}}}\frac{\Phi_{\omega}^{\prime}|_{x\theta}^{x_{\mathrm{r}}}}{\Phi_{\omega}|_{x\theta}^{x_{\mathrm{r}}}},\label{eq:Jannis}
\end{equation}
for the oscillation frequency $\omega$ and\textcolor{brown}{{} }\textcolor{black}{the
boundaries $x_{\left\{ \mathrm{r},\theta\right\} }=\sqrt{2}$$y_{\left\{ \theta,r\right\} }$.
The function $\Phi_{\omega}\left(x\right)=e^{\frac{1}{4}x^{2}}\,U\left(i\omega\tau_{\mathrm{m}}-\frac{1}{2},x\right)$
is defined by parabolic cylinder functions $U$ \cite{Abramowitz74,Lindner01_2934}
and $\Phi_{\omega}^{'}=\partial_{x}\Phi_{\omega}$. $\Phi_{\omega}|_{x\theta}^{x_{\mathrm{r}}}$
is a short-hand notation for $\Phi_{\omega}\left(x_{\mathrm{r}}\right)-\Phi_{\omega}\left(x_{\theta}\right)$.}
We need to multiply the transfer function with the transfer function
of a first-order low-pass filter due to the exponential time course
of our synaptic currents:

\begin{equation}
H_{\mu}\left(\omega\right)=H_{G}\left(\omega\right)\,{\normalcolor }\frac{1}{1+i\omega\tau_{\mathrm{s}}}.
\end{equation}
We then obtain $h_{\mu}$ by an inverse Fourier transform and a Laplace
transform because $\lambda$ is a complex frequency and $\omega$
is real in the present context:

\begin{equation}
\begin{split}h_{\mu}\left(t\right) & =\mathcal{F}^{-1}\left[H_{\mu}\right]\left(t\right)\\
H_{\mu}\left(\lambda\right) & =\mathcal{L}\left[h_{\mu}\right]\left(\lambda\right).
\end{split}
\label{eq:Fourier_Laplace}
\end{equation}
The latter relations imply a replacement $i\omega\to\lambda$ in \prettyref{eq:Jannis}.

\textcolor{brown}{}\textcolor{black}{For completeness we also provide
the term due to the modulation of the variance \cite[Eq 29]{Schuecker15_transferfunction},
cf. also \prettyref{eq:response_kernels},
\begin{align*}
H_{\sigma^{2}}(\omega) & =\frac{1}{\sigma^{2}}\frac{\nu_{0}}{2+i\omega}\,\frac{\Phi_{\omega}^{\prime\prime}|_{x_{\theta}}^{x_{\mathrm{r}}}}{\Phi_{\omega}|_{x_{\theta}}^{x_{\mathrm{r}}}}.
\end{align*}
In the fluctuation-driven regime, $H_{\mu}$ and $H_{\sigma^{2}}$
both have a maximum at vanishing frequency. We compare these two contributions
in \prettyref{fig:transfer_function_fit}B, which shows that $H_{\sigma^{2}}$
can be neglected compared to the former with only making a small error.}\textcolor{brown}{}

\subsection{Model comparison}

\subsubsection{Effective coupling strength}

For the numerical evaluation of the transfer function, we show $H_{0}^{\mathrm{ecs}}=w^{\mathrm{ecs}}/\left(\tau_{\mathrm{m}}JK\right)$
as the dashed line in \prettyref{fig:transfer_function_fit}B, obtained
by calculating analytically the effective coupling strength $w^{\mathrm{ecs}}$
from linear-response theory. The effective coupling strength for a
connection from neuron $j$ with rate $\nu_{j}$ to neuron $i$ with
rate $\nu_{i}$ is defined as \cite[Eqs. A.2 and A.3 (correcting a typo in this previous work)]{Helias13_023002}:

\begin{equation}
\begin{split}w_{ij}^{\mathrm{ecs}} & =\frac{\partial\nu_{i}}{\partial\nu_{j}}\\
 & =\widetilde{\alpha}J_{ij}+\widetilde{\beta}J_{ij}^{2}\\
\text{with}\quad\widetilde{\alpha} & =\sqrt{\pi}\left(\tau_{\mathrm{m}}\nu_{i}\right)^{2}\frac{1}{\sigma_{i}}\left(f\left(y_{\theta}\right)-f\left(y_{\mathrm{r}}\right)\right)\\
\text{and}\quad\widetilde{\beta} & =\sqrt{\pi}\left(\tau_{\mathrm{m}}\nu_{i}\right)^{2}\frac{1}{2\sigma_{i}^{2}}\,\left(f\left(y_{\theta}\right)\,y_{\theta}-f\left(y_{\mathrm{r}}\right)\,y_{\mathrm{r}}\right),
\end{split}
{\normalcolor }\label{eq:effective_coupling_strength}
\end{equation}
where $f$ and $y_{\left\{ \theta,r\right\} }$ are defined as in
\prettyref{eq:siegert}. The dashed line in \prettyref{fig:transfer_function_fit}B
is given by the term $\propto\widetilde{\alpha}$ alone since we also
ignore the small contribution of the variance to the transfer function
($H_{\sigma^{2}}$) of the LIF neuron \cite{Schuecker15_transferfunction}
.

\subsubsection{Linear interpolation}

To compute the derivative $d\lambda/d\alpha$ given in \prettyref{eq:d_lambda_d_alpha},
we use a method for computing the derivative of an implicit function:
If $R\left(\alpha,\lambda\right)=0$, it follows that the derivative

\begin{equation}
\frac{\partial\lambda}{\partial\alpha}=-\frac{\partial R/\partial\alpha}{\partial R/\partial\lambda}=:-\frac{R_{\alpha}}{R_{\lambda}}.\label{eq:derivative_implicit_fct}
\end{equation}
With the characteristic equation for the effective transfer function
\prettyref{eq:comb_char_eq}, we get

\begin{equation}
R\left(\alpha,\lambda\right)=H_{\alpha}\left(\lambda\right)\cdot\mathrm{e}^{-\lambda d}\cdot\widehat{p}\left(k\right)-1=0.
\end{equation}
The partial derivatives of $R$ with respect to $\alpha$ and $\lambda$
are 

\[
\begin{split}R_{\alpha} & =\mathrm{e}^{-\lambda d}\cdot\widehat{p}\left(k\right)\cdot\frac{\partial H_{\alpha}\left(\lambda\right)}{\partial\alpha}\\
 & =\mathrm{e}^{-\lambda d}\cdot\widehat{p}\left(k\right)\cdot\left[H^{\mathrm{s}}\left(\lambda\right)-H^{\mathrm{nf}}\left(\lambda\right)\right],
\end{split}
\]
and

\begin{equation}
\begin{split}R_{\lambda} & =\widehat{p}\left(k\right)\cdot\frac{\partial}{\partial\lambda}\left[H_{\alpha}\left(\lambda\right)\cdot\mathrm{e}^{-\lambda d}\right]\\
 & =\mathrm{e}^{-\lambda d}\cdot\widehat{p}\left(k\right)\cdot\left[\frac{\partial H_{\alpha}\left(\lambda\right)}{\partial\lambda}-d\cdot H_{\alpha}\left(\lambda\right)\right]\\
 & =\mathrm{e}^{-\lambda d}\cdot\widehat{p}\left(k\right)\cdot\left[\alpha\frac{\partial H^{\mathrm{s}}\left(\lambda\right)}{\partial\lambda}+\left(1-\alpha\right)\frac{\partial H^{\mathrm{nf}}\left(\lambda\right)}{\partial\lambda}-d\cdot H_{\alpha}\left(\lambda\right)\right]\\
 & =\mathrm{e}^{-\lambda d}\cdot\widehat{p}\left(k\right)\cdot\left[\alpha H_{\lambda}^{\mathrm{s}}\left(\lambda\right)+\left(1-\alpha\right)\cdot H_{\lambda}^{\mathrm{nf}}\left(\lambda\right)-d\cdot H_{\alpha}\left(\lambda\right)\right].
\end{split}
\label{eq:R_lambda}
\end{equation}

\subsection{Fixing the working point}

\textcolor{brown}{}\textcolor{black}{For the spiking model, we fix
the total input to each neuron in terms of its mean $\mu^{*}$ and
variance $\sigma^{*}$ to given values. To attain a fixed working
point $\left(\mu^{*},\sigma^{*}\right)$, we add to the local contribution
from the recurrently connected network (see \prettyref{eq:mean_var-1})
external excitatory and inhibitory input with Poisson-distributed
interspike interval statistics:
\begin{equation}
\begin{split}\mu^{*} & =\mu+\tau_{\mathrm{m}}J\left(\nu_{\mathrm{E,ext}}-g\nu_{\mathrm{I,ext}}\right)\\
\sigma^{*} & =\sigma+\tau_{\mathrm{m}}J^{2}\left(\nu_{\mathrm{E,ext}}+g^{2}\nu_{\mathrm{I,ext}}\right).
\end{split}
\end{equation}
The excitatory and inhibitory external connection strengths are $J$
and $-gJ$, respectively. The expressions for the excitatory and inhibitory
external rates are:
\begin{equation}
\nu_{\mathrm{E,ext}}=\frac{\widetilde{\sigma}^{2}+g\widetilde{\mu}}{1+g}\qquad\text{and\ensuremath{\qquad\nu_{\mathrm{I,ext}}}\ensuremath{\frac{\widetilde{\sigma}^{2}-\widetilde{\mu}}{g\left(1+g\right)}}}\label{eq:external_rates}
\end{equation}
\[
\text{with}\qquad\widetilde{\mu}=\frac{\mu^{*}-\mu}{\tau_{\mathrm{m}}J}\qquad\text{and}\qquad\widetilde{\sigma}^{2}=\frac{\left(\sigma^{*}\right)^{2}-\sigma^{2}}{\tau_{\mathrm{m}}J^{2}}.
\]
\prettyref{eq:external_rates} corrects a small inconsistency in
a preliminary report of this study \cite{Senk18_arxiv_06046v1} which
used Eq E.1 in \cite{Helias13_023002} to fix the working point. Accordingly,
the values for $\nu_{\mathrm{E,ext}}$ and $\nu_{\mathrm{I,ext}}$
are updated in \prettyref{tab:parameters}, affecting the raster plots
in Figs \ref{fig:rasters_spatiotemporal_patterns} and \ref{fig:transitions}.}

\subsection{Physical units}

The sub-threshold dynamics of the LIF neuron in \prettyref{eq:subthreshold_LIF}
are, without loss of generality, given in scaled units. In this formulation,
$V$, $J$ and $I$ are all quantities with unit Volt. For the parameter-wise
comparison with numerical network simulation (for example using NEST
\cite{Nest2180}), it is useful to consider a description where $I^{'}$
and $J^{'}$ represent electric currents in units of Ampere:

\begin{equation}
\begin{split}\tau_{\mathrm{m}}\frac{\mathrm{\partial}V_{i}^{'}}{\mathrm{\partial}t} & =-\left(V_{i}^{'}-E_{\mathrm{L}}\right)+R_{\mathrm{m}}I{}_{i}^{'}\left(t\right)\\
\tau_{\mathrm{s}}\frac{\mathrm{\partial}I_{i}^{'}}{\mathrm{\partial}t} & =-I_{i}^{'}+\tau_{\mathrm{s}}\sum_{j}J{}_{ij}^{'}s_{j}\left(t-d\right).
\end{split}
\label{eq:lif_dynamics_phy}
\end{equation}

Here, we also introduce a resistive leak reversal potential $E_{\mathrm{L}}$,
and shift threshold and reset potentials $V'_{\theta}=V_{\theta}+E_{\mathrm{L}}$
and $V'_{r}=V_{r}+E_{\mathrm{L}}$, respectively. The membrane time
constant $\tau_{\mathrm{m}}=R_{\mathrm{m}}C_{\mathrm{m}}$ relates
the membrane resistance $R_{\mathrm{m}}$ and capacitance $C_{\mathrm{m}}$.
In units of Ampere, the total current input $I^{'}=I/R_{\mathrm{m}}$
and the synaptic weight amplitude $J^{'}=C_{\mathrm{m}}J/\tau_{\mathrm{s}}$.

\subsection{Network structure and parameters}

\begin{table}[t]
\begin{centering}
\begin{tabular}{|@{\hspace*{1mm}}p{3cm}@{}|@{\hspace*{1mm}}p{9.5cm}|}
\hline 
\multicolumn{2}{|>{\color{white}\columncolor{black}}l|}{\textbf{Model summary}}\tabularnewline
\hline 
\textbf{Populations} & Excitatory ($\mathrm{E}$), inhibitory ($\mathrm{I}$)\tabularnewline
\hline 
\textbf{Topology} & Ring network: Neurons positioned equally spaced on one-dimensional
domain of length $L$; periodic boundary conditions\tabularnewline
\hline 
\textbf{Connectivity} & Random convergent connections with fixed in-degree, distance-dependent
boxcar-shaped spatial profiles realized with cut-off masks\tabularnewline
\hline 
\multicolumn{2}{|>{\color{black}\columncolor{lightgray}}c|}{\textbf{Spiking model}}\tabularnewline
\hline 
\textbf{Neuron model} & Leaky integrate-and-fire (LIF), fixed threshold, absolute refractory
time\tabularnewline
\hline 
\textbf{Synapse model} & Static weights and delays, exponentially shaped postsynaptic currents\tabularnewline
\hline 
\textbf{Input} & Independent fixed-rate Poisson spike trains to all neurons (excitatory
and inhibitory Poisson sources)\tabularnewline
\hline 
\textbf{Measurement} & Spike activity\tabularnewline
\hline 
\multicolumn{2}{|>{\color{black}\columncolor{lightgray}}c|}{\textbf{Rate model}}\tabularnewline
\hline 
\textbf{Neuron model} & Rate neuron with $\mathrm{tanh}$ gain function\tabularnewline
\hline 
\textbf{Synapse model} & Delayed rate connection\tabularnewline
\hline 
\textbf{Input} & -\tabularnewline
\hline 
\textbf{Measurement} & Activity\tabularnewline
\hline 
\end{tabular}
\par\end{centering}
\caption{\textbf{Summary of network models following the guidelines of Nordlie
et al. }\cite{Nordlie09}. Separation between nonlinear spiking and
rate neurons as used in NEST simulations.\label{tab:model_summary}}
\end{table}

We simulate recurrently connected neural networks of one excitatory
and one inhibitory populations each using the neural simulation software
NEST \cite{Gewaltig_07_11204}, using either spiking- or rate-neuron
models. The support for rate neurons in NEST was added as described
in \cite{Hahne17_34}. Figs \ref{tab:model_summary} and \ref{tab:model_description}
provide the complete neuron and network model descriptions and \prettyref{tab:parameters}
summarizes all parameters as used for the network state showing wave
trains (marked by black star in \prettyref{fig:rasters_spatiotemporal_patterns}D,
\prettyref{fig:rate_simulation}D and \prettyref{fig:transitions}C).
Other simulation parameters used to obtain other network states shown
throughout this paper are indicated with a $\varoast$ marker in \prettyref{tab:parameters},
and the changed parameters are given in the corresponding figures.
The same marker always denotes the same parameter combination across
figure panels. The tables distinguish between network properties and
parameters valid for both spiking and rate neuron models and those
specific to only one neuron model. Irrespective of the choice of neuron
model (rate vs. spiking), the neuron parameters are shared between
both neuron populations.

The number of excitatory neurons $N_{\mathrm{E}}$ in our network
is four times larger than the number of inhibitory neurons $N_{\mathrm{I}}$
\cite{Braitenberg01}. \textcolor{black}{All neurons are positioned
on a grid along a one-dimensional path of perimeter $L$ with a space
constant of $\Delta x=L/N_{\mathrm{I}}$. At each grid position $x\in\left[0,L-\Delta x\right]$,
there is one inhibitory neuron and four excitatory neurons. The network
activity in Figs \ref{fig:rasters_spatiotemporal_patterns}, \ref{fig:rate_simulation}
and \ref{fig:transitions} is shown for all inhibitory neurons, but
only for one excitatory neuron at each grid position.} \textcolor{black}{Connections
between neurons are drawn according to a distance-dependent rule with
periodic boundary conditions (a ``ring'' network) using the NEST
Topology module. The number of incoming connections, the in-degree
$K_{\left\{ \mathrm{E,I}\right\} }$, is proportional to the population
size of the presynaptic population, assuming an overall connection
probability of $10\%$. The width of the boxcar-shaped distance-dependent
profile $R_{\left\{ \mathrm{E,I}\right\} }$ depends on the presynaptic
population alone. Within a distance of $R_{\left\{ \mathrm{E,I}\right\} }$
around each postsynaptic neuron, potential presynaptic neurons are
selected at random and connections are established until the prescribed
in-degree is reached. The random component of this connection algorithm
may lead to a slight asymmetry with respect to excitation and inhibition
in the finite-sized network that might cause a small drift visible
for example in stationary Turing patterns as in \prettyref{fig:rate_simulation}B.}
Potentially presynaptic neurons within this distance are picked at
random and connections are established until the fixed in-degree is
reached. Multiple connections between the same pair of neurons termed
multapses are allowed, but self-connections (autapses) are prohibited.

The leaky integrate-and-fire model with exponential postsynaptic currents
is implemented in NEST under the name \texttt{iaf\_psc\_exp}. The
neuron parameters are the same as in the microcircuit model of \cite{Potjans14_785}
with the difference that our membrane time constant $\tau_{\mathrm{m}}$
is half of theirs and that we here omit the refractory period $\tau_{\mathrm{ref}}$,
although our results generalize to a non-zero $\tau_{\mathrm{ref}}$.
An excitatory and an inhibitory Poisson generator provide external
input to all neurons. Their rates $\nu_{\left\{ \mathrm{E,I}\right\} ,\mathrm{ext}}$
are determined according to \prettyref{eq:external_rates} for fixing
the working point $\left(\mu,\sigma\right)$.

The dynamics of rate-based units in NEST is specified as stochastic
differential equations using the It\^{o} convention \cite{Hahne17_34},
except that we here set the stochasticity (the variance of the input)
to zero. We use the neuron model \texttt{tanh\_ipn}, that employs
a hyperbolic tangent as a gain function.

Simulations run for a simulation time $T_{\mathrm{sim}}$ with a temporal
resolution of $dt$. During rate simulations, the instantaneous rate
is recorded once at each time step $dt$. Our raster plots from simulations
of the spiking model and the image plots from simulation of the rate
model show the network activity from all simulated neurons after a
start-up transient $T_{\mathrm{trans}}.$

\subsection{Software and implementation}

Spiking- and rate-neuron network simulations were implemented in NEST
v2.18.0 \cite{Nest2180}, and Python v3.6.9. Post-processing and plotting
relied on Python with NumPy v1.16.4, SciPy v1.2.1, and Matplotlib
v3.0.2.

\begin{table}[H]
\begin{centering}
\begin{tabular}{|@{\hspace*{1mm}}p{3cm}@{}|@{\hspace*{1mm}}p{9.5cm}|}
\hline 
\multicolumn{2}{|>{\color{white}\columncolor{black}}l|}{\textbf{Network models}}\tabularnewline
\hline 
\textbf{Distance-dependent\linebreak connectivity} & Neural units $j\in X$ at location $x_{j}$ and $i\in Y$ at $x_{i}$
in pre- and postsynaptic populations $X$ and $Y$, respectively.\newline
Displacement between units $i$ and $j$:\begin{itemize} \item[]$r_{ij}=x_{i}-x_{j}$\end{itemize}
Boxcar-shaped spatial profile with width $R$ and Heaviside function
$\Theta$:\begin{itemize} \item[]$p\left(r_{ij}\right)=\frac{1}{2R}\Theta\left(R-\left|r_{ij}\right|\right)$\end{itemize} \tabularnewline
\hline 
\multicolumn{2}{|>{\color{black}\columncolor{lightgray}}c|}{\textbf{Spiking model}}\tabularnewline
\hline 
\textbf{Subthreshold\linebreak dynamics} & If $t>t^{*}+\tau_{\mathrm{ref}}$\begin{itemize}\item[]$\frac{\mathrm{\partial}V}{\partial t}=-\frac{V-E_{\mathrm{L}}}{\tau_{\mathrm{m}}}+\frac{I_{\mathrm{syn}}\left(t\right)}{C_{\mathrm{m}}}$\item[]$I_{\mathrm{syn}}\left(t\right)=\sum_{j}J_{j}I_{\mathrm{PSC}}\left(t-t_{j}^{*}-d\right)$\item[]
with connection strength $J_{j}$, presynaptic spike time $t_{j}^{*}$
and conduction delay $d$ \item[]$I_{\mathrm{PSC}}\left(t\right)=\mathrm{e}^{-t/\tau_{\mathrm{s}}}\Theta\left(t\right)$
with Heaviside function $\Theta$\end{itemize} else \begin{itemize}
\item[] $V\left(t\right)=V_{r}$\end{itemize}\tabularnewline
\hline 
\textbf{Spiking} & If $V\left(t-\right)<V_{\theta}\wedge V\left(t+\right)\geq V_{\theta}$\begin{enumerate}\item
set $t^{*}=t$ \item emit spike with timestamp $t^{*}$ \item reset
$V\left(t\right)=V_{\mathrm{r}}$\end{enumerate}\tabularnewline
\hline 
\multicolumn{2}{|>{\color{black}\columncolor{lightgray}}c|}{\textbf{Rate model}}\tabularnewline
\hline 
\textbf{Differential\linebreak equation} & \textcolor{black}{$\tau\frac{\mathrm{\partial}u}{\mathrm{\partial}t}\left(t\right)=-u\left(t\right)+\sum_{j=1}w_{j}\psi\left(u_{j}\left(t-d\right)\right)$
with the nonlinearity $\psi\left(u\right)=\tanh\left(u\right)$}\tabularnewline
\hline 
\end{tabular}
\par\end{centering}
\caption{\textbf{Description of network models.} Separation between nonlinear
spiking and rate neurons as used in NEST simulations. \label{tab:model_description}}
\end{table}

\begin{table}[H]
\begin{centering}
\begin{tabular}{|@{\hspace*{1mm}}p{2cm}@{}|@{\hspace*{1mm}}p{3cm}|@{\hspace*{1mm}}p{7.5cm}|}
\hline 
\multicolumn{3}{|>{\color{white}\columncolor{black}}l|}{\textbf{A: Global simulation parameters}}\tabularnewline
\hline 
\textbf{Symbol} & \textbf{Value} & \textbf{Description}\tabularnewline
\hline 
$T_{\mathrm{sim}}$ & $\unit[350]{ms}$ & Simulation duration\tabularnewline
\hline 
$T_{\mathrm{trans}}$ & $\unit[150]{ms}$ & Start-up transient\tabularnewline
\hline 
$dt$ & $\unit[0.1]{ms}$ & Temporal resolution\tabularnewline
\hline 
\end{tabular}
\par\end{centering}
\begin{centering}
\begin{tabular}{|@{\hspace*{1mm}}p{2cm}@{}|@{\hspace*{1mm}}p{3cm}|@{\hspace*{1mm}}p{7.5cm}|}
\hline 
\multicolumn{3}{|>{\color{white}\columncolor{black}}l|}{\textbf{B: Populations and external input}}\tabularnewline
\hline 
\textbf{Symbol} & \textbf{Value} & \textbf{Description}\tabularnewline
\hline 
$N_{\mathrm{E}}$ & $4,000$ & Population size of excitatory neurons\tabularnewline
\hline 
$N_{\mathrm{I}}$ & $1,000$ & Population size of inhibitory neurons\tabularnewline
\hline 
$L$ & $\unit[1]{mm}$ & Domain length\tabularnewline
\hline 
\multicolumn{3}{|>{\color{black}\columncolor{lightgray}}c|}{\textbf{Spiking model}}\tabularnewline
\hline 
$\mu^{*}$ & $\unit[10]{mV}$ & Mean input relative to resting potential\tabularnewline
\hline 
$\sigma^{*}$ & $\unit[10]{mV}$ & Variance of input relative to resting potential\tabularnewline
\hline 
$\nu_{\mathrm{E,ext}}$ & $\unit[96,463]{Hz}$ & $\varoast$ Excitatory external rate (by fixing working point)\tabularnewline
\hline 
$\nu_{\mathrm{I,ext}}$ & $\unit[15,958]{Hz}$ & $\varoast$ Inhibitory external rate (by fixing working point)\tabularnewline
\hline 
\end{tabular}
\par\end{centering}
\begin{centering}
\begin{tabular}{|@{\hspace*{1mm}}p{2cm}@{}|@{\hspace*{1mm}}p{3cm}|@{\hspace*{1mm}}p{7.5cm}|}
\hline 
\multicolumn{3}{|>{\color{white}\columncolor{black}}l|}{\textbf{C: Connection parameters}}\tabularnewline
\hline 
\textbf{Symbol} & \textbf{Value} & \textbf{Description}\tabularnewline
\hline 
$R_{\mathrm{E}}$ & $\unit[0.2]{mm}$ & $\varoast$ Profile width of excitatory neurons\tabularnewline
\hline 
$R_{\mathrm{I}}$ & $\unit[0.07]{mm}$ & $\varoast$ Profile width of inhibitory neurons\tabularnewline
\hline 
$d$ & $\unit[3]{ms}$ & $\varoast$ Delay\tabularnewline
\hline 
\multicolumn{3}{|>{\color{black}\columncolor{lightgray}}c|}{\textbf{Spiking model}}\tabularnewline
\hline 
$K_{\mathrm{E}}$ & $400$ & In-degree from excitatory neurons\tabularnewline
\hline 
$\gamma$ & $0.25$ & Relative in-degree, $\gamma=K_{\mathrm{I}}/K_{\mathrm{E}}$\tabularnewline
\hline 
$J_{\mathrm{E}}^{'}$ & $\unit[87.8]{pA}$ & $\varoast$ Reference synaptic strength\tabularnewline
\hline 
$g$ & $5$ & $\varoast$ Relative synaptic strength, $g=-J_{\mathrm{I}}/J_{\mathrm{E}}$\tabularnewline
\hline 
\multicolumn{3}{|>{\color{black}\columncolor{lightgray}}c|}{\textbf{Rate model}}\tabularnewline
\hline 
$w_{\mathrm{E}}$ & $2.73$ & $\varoast$ Excitatory weight (by parameter mapping)\tabularnewline
\hline 
$w_{\mathrm{I}}$ & $-3.42$ & $\varoast$ Inhibitory weight (by parameter mapping)\tabularnewline
\hline 
\end{tabular}
\par\end{centering}
\begin{centering}
\begin{tabular}{|@{\hspace*{1mm}}p{2cm}@{}|@{\hspace*{1mm}}p{3cm}|@{\hspace*{1mm}}p{7.5cm}|}
\hline 
\multicolumn{3}{|>{\color{white}\columncolor{black}}l|}{\textbf{D: Neuron model}}\tabularnewline
\hline 
\textbf{Symbol} & \textbf{Value} & \textbf{Description}\tabularnewline
\hline 
\multicolumn{3}{|>{\color{black}\columncolor{lightgray}}c|}{\textbf{Spiking model}}\tabularnewline
\hline 
$C_{\mathrm{m}}$ & $\unit[250]{pF}$ & Membrane capacitance\tabularnewline
\hline 
$\tau_{\mathrm{m}}$ & $\unit[5]{ms}$ & Membrane time constant\tabularnewline
\hline 
$E_{\mathrm{L}}$ & $\unit[-65]{mV}$ & Resting potential\tabularnewline
\hline 
$V_{\theta}$ & $\unit[-50]{mV}$ & Firing threshold\tabularnewline
\hline 
$V_{\mathrm{r}}$ & $\unit[-65]{mV}$ & Reset potential\tabularnewline
\hline 
$\tau_{\mathrm{ref}}$ & $\unit[0]{ms}$ & Absolute refractory period\tabularnewline
\hline 
$\tau_{\mathrm{\mathrm{s}}}$ & $\unit[0.5]{ms}$ & Postsynaptic current time constant\tabularnewline
\hline 
\multicolumn{3}{|>{\color{black}\columncolor{lightgray}}c|}{\textbf{Rate model}}\tabularnewline
\hline 
$\tau$ & $\unit[1.94]{ms}$ & Time constant (by parameter mapping)\tabularnewline
\hline 
\end{tabular}
\par\end{centering}
\centering{}\caption{\textbf{Simulation and network parameters. }Parameters according to
setting for wave trains as shown in \prettyref{fig:rasters_spatiotemporal_patterns}D,
\prettyref{fig:rate_simulation}D and \prettyref{fig:transitions}C
(black star marker). Deviant parameters are given in the captions
of the respective figures and indicated by different markers. \label{tab:parameters}}
\end{table}

\section*{Acknowledgments}

The authors would like to thank the whole INM-6 for fruitful discussions;
in particular Michael Denker and Sonja Grün for sharing their experience
on wave-like activity in experimentally recorded brain activity, and
David Dahmen, Hannah Bos and other colleagues from the NEST community
(\href{http://www.nest-simulator.org}{http://www.nest-simulator.org}).

\section*{Funding}

This project has received funding from the Helmholtz association:
portfolio Supercomputing and Modeling for the Human Brain (SMHB),
young investigator group VH-NG-1028; the European Union's Horizon
2020 research and innovation programme under grant agreement No 720270
(HBP SGA1), the German Research Foundation (DFG; grant DI 1721/3-1
{[}KFO219-TP9{]}), the ERS grant \textquotedbl Facing the multi-scale
problem in neuroscience\textquotedbl{} of the RWTH Aachen University,
and the Research Council of Norway (NFR) through COBRA (grant No 250128).
The funders had no role in study design, data collection and analysis,
decision to publish, or preparation of the manuscript.

\end{document}